# TO BETTER UNDERSTAND THE FORMATION OF SHORT-CHAIN ACIDS IN COMBUSTION SYSTEMS

F. BATTIN-LECLERC[1*], A.A. KONNOV[2], J.L. JAFFREZO[3] AND M. LEGRAND[3]

[1]Département de Chimie-Physique des Réactions, CNRS-INPL, Nancy, France

[2]Vrije Universiteit Brussel, Brussels, Belgium

[3]Laboratoire de Glaciologie et Géophysique de l'Environnement, CNRS, St Martin d'Hères, France

[*] Corresponding author: DCPR-ENSIC, BP 20541, 1 rue Grandville, BP 20451, 54001 NANCY Cedex, France

E-mail : Frederique.battin-leclerc@ensic.inpl-nancy.fr ; Tel.: 33 3 83 17 51 25 , Fax : 33 3 83 37 81 20

**Abstract**

While their presence at the outlet of IC engines has been attested, the formation of short-chain monocarboxylic acids, formic (HCOOH), acetic ($CH_3COOH$), propionic ($C_2H_5COOH$) or acrylic ($C_2H_3COOH$)) acids has very rarely been reported in laboratory combustion systems. In order to better understand their potential formation, detailed kinetic mechanisms have been proposed and tested. A first model has been used to simulate lean (equivalence ratios from 0.9 to 0.48) laminar premixed flames of propane stabilized at atmospheric pressure. It was found that amounts up to 40 ppm of formic acid, 25 ppm of acetic acid and 1 ppm of $C_3$ acids, mainly acrylic acid, can be formed. A quite acceptable agreement has been obtained with the scarce results from the literature concerning oxygenated compounds, including aldehydes and acids. A reaction pathways analysis demonstrated that each acid is mainly derived from the aldehyde of similar structure, with a dominant role of OH• radicals. Based on this first one, a second model has allowed us to simulate a flame of propane doped by toluene and to show, as it was experimentally observed, an enhancement of the formation of $C_3$ acid, which could be due to the addition of OH• radicals to cyclopentadienone. A third model has been proposed to qualitatively explain the formation of acids during the pre-ignition phase (temperatures below 1100 K) in an HCCI (Homogeneous Charge Compression Ignition) engine alimented by a n-heptane/iso-octane mixture (equivalence ratio of 0.3). Noticeable amounts of monocarboxylic acids could derive from the secondary reactions of ketones or cyclic ethers, which are important products of the oxidation of alkanes at low temperature. These amounts remain too low compared to what is actually observed at the outlet of engines.

## Introduction

As shown by the review of Chebbi and Carlier (1996), in addition to sulfuric and nitric acids, monocarboxylic acids (formic and acetic acids, especially) can contribute to the acidity of gas and aqueous phases of the atmosphere in urban, as well as in remote regions. Natural and anthropogenic sources of monocarboxylic acids have been proposed, but the nature of the individual sources and their relative importance are not established yet (Legrand et al., 2003). While numerous olefin-ozone reactions can produce formic and acetic acids (Calvert and Stockwell, 1983), it is of importance to better understand the potential direct emissions of monocarboxylic acids from combustion phenomena, from automotive engines especially.

Detailed analyses of gas-phase pollutants emitted from the exhaust have been published for diesel (Zervas et al., 2001a) and gasoline (Zervas et al., 2001b, 2002, 2003) engines. These papers describe the presence of noticeable amounts of monocarboxylic acids. These acids are mainly formic, acetic and propionic acids; the formation of this last one being strongly linked with the presence of aromatic compounds in fuel. In a gasoline engine, the emission of acids corresponds to 4 to 27% of the total amount of emitted hydrocarbons and is larger by a factor of 1.3 to 10 than that of aldehydes; the major compounds being acetic and propionic acids (Zervas et al., 2001b, 2002).

Despite that, acids are very minor oxidation products in most laboratory experimental gas-phase apparatuses and there are very few studies showing their formation during combustion or kinetic model proposed to explain their formation and consumption pathways. Zervas (2005) has analyzed the formation of acids in laminar premixed flames of propane, iso-octane and iso-octane/toluene at atmospheric pressure, for equivalence ratios from 0.8 to 1.2. Curran et al. (2000) have found the presence of formic acid during the oxidation of dimethylether in a flow reactor operated over an initial temperature range of 550–850 K, in the pressure range 12–18 atm, for equivalence ratios from 0.7 to 4.2, and proposed some reaction channels to explain its formation.

The purpose of this paper is first to model[1] the formation of light acids under flame conditions and then to better understand the possible reaction channels which could be responsible for their presence at lower temperature, during the pre-ignition phase in HCCI engines.

## Modeling of the formation of acids under flame conditions

The work of Zervas (2005) seems to be the only one showing the presence of light acids under laminar flame conditions; unfortunately, all the data needed for modeling was not provided. By chance, a very detailed study of the temperatures and species profiles in a laminar premixed flame of propane at atmospheric pressure, for equivalence ratios from 0.48 to 0.9, has been recently published by Biet et al. (2005). As the two studies have been performed for some common equivalence ratios, we attempted here to implement a new kinetic mechanism of the formation of acids in a model able to well reproduce the results of Biet et al. (2005) and to use this global model to simulate the relative formation of monocarboxylic acids in the flame of Zervas (2005). Zervas has also observed that the nature of the initial alkane does not much influence the relative formation of acids, but that the presence of toluene increases the formation of $C_3$ acid. In order to find the chemical reaction responsible for this enhancement, we have added to our first model a submechanism for the oxidation of toluene and run simulations for a propane flame seeded with toluene.

### *Development of a kinetic model to reproduce the formation of acids in a propane flame*

A mechanism has been written to reproduce the formation and consumption of formic, acetic, propionic and acrylic acids under flame conditions; i.e. the low-temperature reactions of peroxy radicals have not been considered.

[1] All the models described in this paper are available on request (Frederique.battin-leclerc@ensic.inpl-nancy.fr).

This new mechanism was based on the kinetic scheme (about 800 reactions) of the combustion of propane recently proposed to model laminar premixed flames of methane doped with allene or propyne (Gueniche et al., 2006). This $C_3$-$C_4$ reaction base, which was described in details in a previous paper (Fournet et al., 1999), was built from a review of the recent literature and is an extension of our previous $C_0$-$C_2$ reaction base (Barbé et al., 1995). The $C_3$-$C_4$ reaction base includes reactions involving all the $C_3$ no-oxygenated species and acrolein, as well as some reactions for selected $C_4$ and $C_5$ compounds and the formation of benzene. In this reaction base, pressure-dependent rate constants follow the formalism proposed by Troe (1974) and collisional efficiency coefficients have been included.

To extend this model in order to describe the formation and consumption of acids, the relevant sub-mechanism was adopted from the kinetic scheme proposed by Konnov (2006). The number of species was increased by 31 and the number of new reactions was 197. The formulae and the thermochemical data of the new species are presented in Table I. The complete set of added reactions is given in Table II.

**TABLES I AND II**

Thermodynamic data of the following species: HOCO•, HCOO• (formyl radical), HCOOH (formic acid), HCCOH, CH2CHOH (vinyl alcohol), CHOCHO, $CH_3$CO2H (acetic acid) and $C_2H_5$CO• (propionaldehyde radical) were taken from the latest database of Burcat and Ruscic (2006). HiTempThermo database (Allendorf, 2006) was used to obtain the thermochemical data for HCOH• (hydroxymethylene), $HOCH_2O$•, $HOCH_2OO$•, $C_2H_2OH$• (HOCH=CH•), $H_2CCOH$•, •$CH_2CH_2OOH$, •$CH_3CHOOH$ and other species were Estimated using the software THERGAS (Muller et al., 1995), which is based on the additivity methods proposed by Benson (1976).

Whenever possible, rate constants have been adopted from the literature and reviews, for instance, of Atkinson et al. (1997, 2005); 30 rate constants have been thus found. In many cases, however, rate constants were Estimated assuming same reactivity of similar species. For example, the reactivity of $CH_2(OH)_2$, $HOCH_2OO$•, $HOCH_2OOH$ and $H_2CCOH$• were assumed analogous to $CH_3OH$, $CH_3O_2$, $CH_3O_2H$ and $C_2H_2OH$•, respectively. Strictly speaking this approach could be more or less valid for true homologues; remote similarity could be misleading due to resonance stabilization of radicals, significant modifications of the bond dissociation energy, etc. On the other hand, this is probably the only way for exploratory modelling of the formation of acids, which are usually not included in the contemporary combustion mechanisms.

A few modifications of the sub-mechanism adopted from the Konnov (2006) kinetic scheme were made. The rate constant of reaction (12) of addition of OH• radicals to formaldehyde is taken as 0.3% that of the H-abstraction, which is in agreement with the value of 4% proposed by Sivakumaran et al. (2003) as an upper limit for this branching ratio at 298 K. This experimental study of the reaction between OH• radicals with formaldehyde by laser photolysis led Sivakumaran et al. (2003) to think that the rate constant of the addition was very small as no hydrogen atoms could be observed by resonance fluorescence. As discussed by Taylor et al. (2006), the addition of OH• radicals to acetaldehyde giving methyl radicals and formic acid or H-atoms and acetic acid could have been envisaged, but has been shown to be negligible compared to the two possible H-abstractions. The rate constant of the combinations between OH• and carbonyl radicals was taken as that proposed by Allara et al. (1980) for H-atoms. The reactions of propionic and acrylic acids have been deduced from those proposed for acetic acid.

*Validation of the model by modeling the evolution of the main species in a propane flame*

All the simulations have been performed using PREMIX code of the CHEMKIN II collection (Kee et al., 1993) with Estimated transport coefficients and temperature profiles taken from the experiments. Comparisons between experimental results of Biet et al. (2005) and simulations of a laminar premixed propane flame are shown in fig. 1 for $\phi$= 0.9 and in fig. 2 for $\phi$= 0.48.

**FIGURES 1 AND 2**

As it can be seen in fig. 1a and 2a, a correct agreement is obtained for the consumption of propane and oxygen and the formation of carbon dioxide for both equivalence ratios. Figure 1b and 1c show that the formation of water, hydrogen, carbon monoxide and $C_2$ compounds are also well reproduced by the model at $\phi$= 0.9. Similarly good agreement is obtained for $\phi$= 0.48, as it can be seen in fig. 2b and 2c. Figures 1d and 2d display the profiles of aldehydes showing that the modelling of formaldehyde and acetaldehyde is also acceptable. These figures also present our prediction for propionaldehyde and acrolein, which are in agreement with the limit of detection of 100 ppm given by Biet et al. (2005).

It is worth noting that the decrease in equivalence ratio does not change markedly the formation of carbon monoxide and aldehydes. That was well explained by Biet et al. (2005) by defining an "effective equivalence ratio" based on the proportion of oxygen consumed in the flame and showing that it was always close to 1 when oxygen was in excess.

*Simulation of the formation of short-chain carboxylic acids in a propane flame*

Figure 3 presents the simulated profiles of acids obtained under the same conditions as the flame of Biet et al. (2005) using the mechanism above described. This figure shows that small amounts of acids are actually produced in this flame, the major ones being formic and acetic acids. Amongst $C_3$ acids, the main one is acrylic acid. That is directly related to the activation energy for the beta-scission involving the breaking of the C-CO bond for $CH_2$=CH-CO• radicals (Ea = 34.0 kcal/mol (Marinov et al., 1996), ΔHr (298K) = 26.2 kcal/mol) and for CH3-CH2-CO• radicals (Ea = 14.4 kcal/mol (reaction 174 in Table II), ΔHr (298K) = 11.9 kcal/mol); everywhere in the text, carbonyl radicals (R-C•(=O) are noted RCO•. The unsaturated oxygenated radical is less easily decomposed than the saturated one to give carbon monoxide (by a factor $6x10^{-3}$ at 1400 K), its concentration is then larger, which favours its combination with OH• radicals.

**FIGURE 3**

Figure 3 also shows that the formation of acids in these lean flames does not change much when decreasing the equivalence ratio; only the maximum mole fractions of formic and acrylic acids are slightly increased. That is in agreement with what was noticed by Biet et al. (2005) for the formation of aldehydes, but in disagreement with the experimental results of Zervas (2005) showing an increase of about a factor 2 of the formation of formic and acetic acids when varying the equivalence ratio from 1.17 to 0.77.

Figure 4 displays a comparison between the computed proportions (the ratio between the mole fraction of each acid and the sum of their mole fractions) of acids and the experimental ones measured by Zervas et al. (2005) for an equivalence ratio of 0.9. In Zervas's work, organic acids were collected in deionised water and analysed by two methods: ionic chromatography/conductimetry detection for formic acid and gas chromatography/flame ionisation detection for other acids. Details about the collection, the analysis and the calibration can be found in a previous paper of Zervas et al. (1999). The analysis by gas chromatography/flame ionisation detection has been tested for nine aliphatic acids from $C_2$ to $C_7$, but not for acrylic acid. It is therefore difficult to know if the two $C_3$ acids could have been separated under these conditions. Figure 4 shows that, apart from the disagreement concerning the nature of the $C_3$ acid, the proportion between $C_1$, $C_2$ and $C_3$ acids is well captured by the model.

**FIGURE 4**

Figure 5 presents the main pathways of formation of the four monocarboxylic acids studied for an equivalence ratio of 0.9. It can be seen that these acids are mainly derived from the aldehyde having the same structure.

- Formic acid is 99% obtained from the addition of OH• radicals to formaldehyde, followed by the elimination of a hydrogen atom.
- Acetic acid is mainly formed from the combination of OH• and $CH_3CO$• radicals. These last radicals are produced for 40% from reactions of acetaldehyde (directly by abstraction of the carboxylic H-atom or via the formation of •$CH_2CHO$ radicals followed by a beta-scission and the addition of H-atoms to ketene, $CH_2$=C=O), but also for 40% from propene and for 20% from vinyl radicals, which are mainly obtained from reactions of ethylene. Propene and vinyl radical can react with O-atoms to give ketene. The reactions of ketene with H-atoms leading to $CH_3CO$• radicals are responsible for 70% of the formation of these radicals.
- Propionic acid is derived from $C_2H_5CO$• radicals by combination with OH• radicals. $C_2H_5CO$• radicals are produced for 60% from propionaldehyde by H-abstraction by small radicals and for 40% from reactions between methyl radicals and ketene.
- Acrylic acid is only produced from acrolein by H-abstraction by small radicals, followed by a combination of the obtained carbonyl ($C_2H_3CO$•) radicals with OH• radicals.

**FIGURE 5**

*Simulation of the formation of short-chain carboxylic acids in a propane flame doped by toluene*

The sub-mechanisms (about 360 reactions) for the oxidation of benzene and toluene developed by Da Costa et al. (2003) and Bounaceur et al. (2005), respectively, have been added to the above described model. As we are not aware of experimental results which could have been of help for validation, simulations have been run under the same conditions as for the flame of propane of Biet et al. (2005) at $\phi$=0.9, but with an addition of 10% toluene (relative to the mole fraction of propane). The computed proportions of acids are mostly unchanged, the amounts of each compound being slightly lower, i.e. the maximum mole fraction is 30 ppm for formic acid, 18 ppm for acetic acid and 0.2 ppm for acrylic acid. This decrease is due to a lower concentration of OH• radicals in the flame in presence of toluene, which easily leads to the formation of resonance stabilised benzyl radicals.

Figure 6 presents the computed profiles of the main formed aromatic species, toluene, benzene, cyclopentadiene, phenol, benzaldehyde, cyclopentadienone, ethylbenzene, styrene, cresols and benzoquinone. Several secondary reactions of these compounds leading to $CH_2$=CH-CO• radicals have been tested, but the only channel leading to a strong increase in the formation of $C_3$ acids was the following addition of OH• radicals to cyclopentadienone (198):

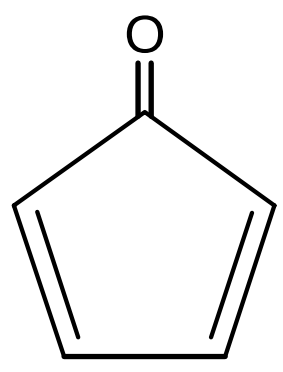

+ •OH ⟶ $C_2H_3COOH$ + C2H•

(198)

with an A-factor of $2 \times 10^{12}$ $cm^3 mol^{-1} s^{-1}$ and an activation energy of -1 kcal/mol, which are usual rate parameters for an addition to a double bond. This reaction involves the opening of the cycle and a rearrangement involving a H-transfer. The reaction between OH• and cyclopentadienone has never been experimentally investigated. During their study of the oxidation of benzene, Alzueta et al. (2000) have written the abstraction of H-atoms (199):

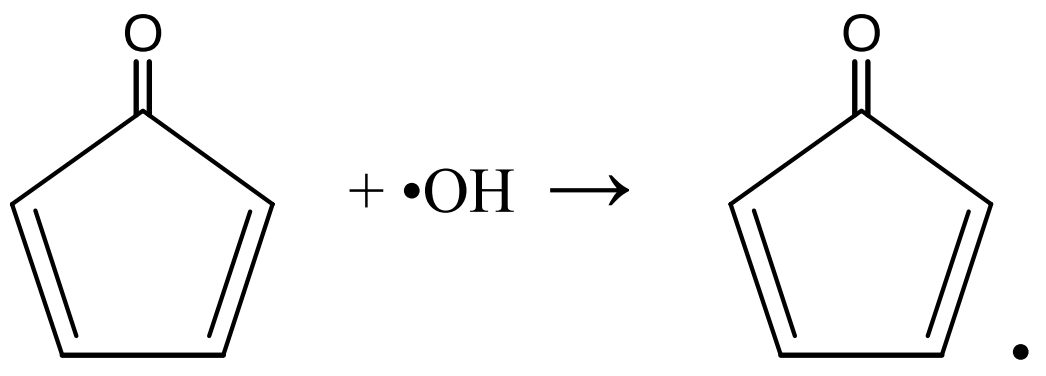

$+ H_2O$ (199)

which is also considered in our mechanism. The addition of OH• radicals to acetone leading to the formation of acetic acid and methyl radicals has been proposed as a very minor channel of the reaction between OH• radicals and acetone. Talukdar et al. (2003), Vandenberk et al. (2002) and Tyndall et al. (2002) have reported an upper limit of 1%, 5% and 10 %, respectively. The addition of OH• radicals to other linear ketones (propanone, butanone, pentanone) has been found negligible by Tranter and Walker (2001), as the formation of acids was not observed.

**FIGURE 6**

Figure 7 presents the simulated profiles of acids obtained with a mechanism including reaction (198) under the same conditions as the flame of propane of Biet et al. (2005) at $\phi$=0.9, but with an addition of 10% toluene (relative to the mole fraction of propane), and shows an important increase of the formation of acrylic acid. The maximum mole fraction of this $C_3$ acid is thus multiplied by a factor above ten by the addition of toluene and is now the major acid in the flame. In the iso-octane/toluene flame studied by Zervas (2005), the $C_3$ acid is also the major one, but is identified as propionic acid.

**FIGURE 7**

## Modeling of the formation of acids during the pre-ignition phase in an HCCI engines

As shown in the previous sections, under flame conditions, the predicted amounts of acids are about forty times lower than that of aldehydes, explaining why the formation of acids is usually not reported in flame experiments. If the emissions of acids from engines are really larger than those of aldehydes, as reported by Zervas et al. (2001b) at the exhaust of gasoline engines, low temperature reactions should be of particular importance to explain this observation. In order to better understand these reactions, simulations have been run under the HCCI conditions used by Dubreuil et al. (2007) using the SENKIN code (Kee et al., 1993) to compute the evolution of a 25% n-heptane/75% iso-octane mixture in air ($C_7H_{16}$ : 0,413% ; $C_8H_{18}$ : 0,1377% ; $O_2$ : 20,9% ; $N_2$ : 78,55%, $\phi = 0.3$ ) in a variable volume, zero-dimensional adiabatic single zone reactor. The kinetic model (around 7800 reactions) has been automatically generated by the EXGAS software for the oxidation of a n-heptane/iso-octane/1-heptene mixture. The rules of generation have been described by Warth et al. (1998) and Buda et al. (2005) for alkanes and by Touchard et al. (2005) for alkenes. Low-temperature reactions have been considered, including all the reactions of peroxy, hydroperoxy and peroxyhydroperoxy radicals. The need to consider a comprehensive model for the oxidation of 1-heptene, a secondary product of n-heptane, will be seen further in the text.

Single-zone modelling allows a satisfactory prediction of ignition delay times as shown by Dubreuil et al. (2007), but not at all a good modelling of the composition of exhaust gases, the final products under lean conditions being only carbon dioxide and water. Even the formation of carbon monoxide is underpredicted by a factor $10^5$. However, assuming that the temperature gradients in the combustion chamber are weak when the temperature rise is still slow (see fig. 8), these computations could give some clues about the species evolution during the pre-ignition phase, as displayed in fig. 8 for the main $C_1$-$C_3$ and $C_7$-$C_8$ oxygenated species. This figure shows that the evolution of the mole fraction of each of these species presents a “plateau” during the pre-ignition zone, i.e. for residence times between $17.5 \times 10^{-3}$ and $18.5 \times 10^{-3}$ s. Assuming that the presence of oxygenated organic species in the exhaust gases is due to the existence of cold zones of the combustion chamber (e. g. near the walls or in crevices) (Aceves et al., 2004), in which the ignition does not reach its final stage, their repartition is probably close to what is observed during this “plateau”. While the combustion phenomena are different in HCCI and spark-ignited engines (homogeneous auto-ignition in the fist one and premixed flame propagation in the second one), the chemistry occurring in the cold zones is probably similar in both systems. Figure 9 shows that the mole fractions of acetaldehyde, acrolein, propionaldehyde and methanol relative to that of formaldehyde and obtained by our simulation under the conditions of fig. 8 at a residence time of $18.5x10^{-3}$ s are in good agreement with those measured by Zervas et al. (2002) in a gasoline engine alimented by a fuel rich in iso-octane at an equivalence ratio of 1. This agreement, which is unexpectedly satisfactory according to the roughness of the hypotheses made, the difference of equivalence ratios and the possibility of reaction in the exhaust line, led us to think that our simulation was not unreasonable and that it could be used to investigate the ways of formation of acids in engines.

**FIGURES 8-9**

Figure 10 presents the evolution obtained for the different $C_1$-$C_3$ acids and formaldehyde using the five following models:

- Model 1 is obtained by just adding the sub-mechanism for the formation and consumption of acids previously described for flames simulation to the model of the low-temperature oxidation of a n-heptane/iso-octane/1-heptene mixture. The formation of acids is very small and occurs mainly during the autoignition, when the concentration of OH• radicals is large enough; the main compound is formic acid, but its concentration in the pre-ignition zone (at a residence time of $18.5x10^{-3}$s) is about 1500 times less than that of formaldehyde. The concentrations of acetic and propionic acids are still $10^5$ times lower than that of formic acid.
- Model 2 also contains additional reactions for $CH_3CO(OO•)$ radicals, which are obtained by addition to oxygen molecules of $CH_3CO•$ radicals, which derive from acetaldehyde. The reactions (200) and (201) proposed by Tomas et al. (2001) have been included:

$$CH_3CO(OO\bullet) + HO_2\bullet \longrightarrow CH_3CO(OOH) + O_2 \quad (200)$$

$$CH_3CO(OO\bullet) + HO_2\bullet \longrightarrow CH_3CO(OH) + O_2 + O \quad (201)$$

with an A-factor of $3.8x10^{11}$ $cm^3mol^{-1}s^{-1}$ for reaction (200) and of $7.7x10^{10}$ $cm^3mol^{-1}s^{-1}$ for reaction (201) and an activation energy of –1.8 kcal/mol for both reactions, as determined by these authors. As reaction (200) involves a formation of acid from a radical deriving from acetaldehyde, it was found promising. The addition of these reactions has an effect on the production of acetic acid in the pre-ignition zone, but, while the concentration of this acid has been multiplied by a factor almost 100 at a residence time of $18.5\times10^{-3}$s, this effect is not strong enough. Around 1000 K, the temperature during the preignition zone, the decomposition of $CH_3CO$• radicals is about 1000 times faster than the addition to oxygen molecules.

- Model 3 includes the decomposition (202) of the hydroxyhydroperoxy radicals, which are obtained from 1-heptene, a secondary product of the oxidation of n-heptane, by addition of OH• radicals, followed by an addition to oxygen molecules and isomerizations:

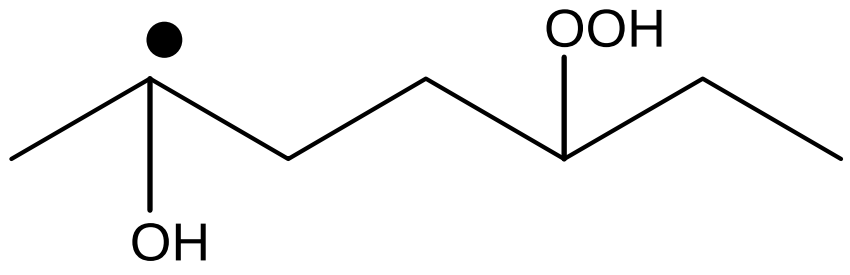

$\longrightarrow$ OH• + $CH_3CO(OH) + C_2H_4 + C_3H_6$ (202)

The globalized reaction (202) is written instead of reaction (203), which involves the formation of a hydroxy cyclic ether, a species never experimentally observed during the oxidation of alkenes (Touchard et al., 2005):

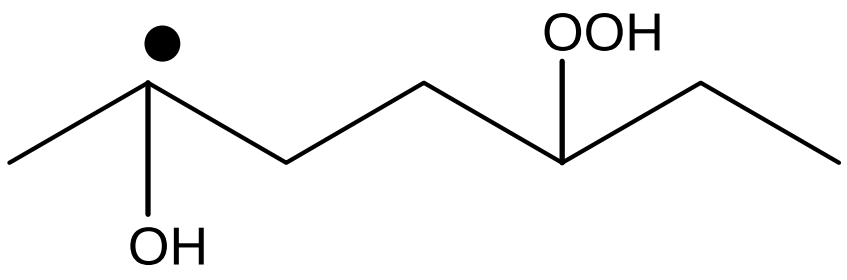

$\longrightarrow$ OH• +

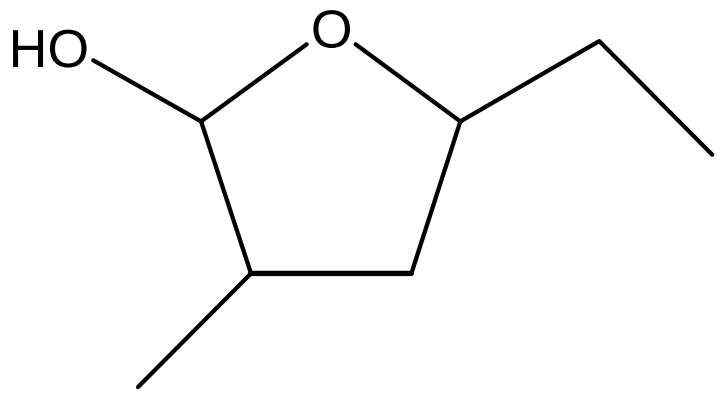

(203)

Considering that both exothermic reactions occur through the same transition state, the same rate parameters have been assumed, an A-factor of $3.6\times10^{9}$ $s^{-1}$ and an activation energy of 7 kcal/mol, as proposed by Touchard et al. (2005). The addition of this reaction has a clear influence on the production of acetic acid in the pre-ignition zone, its concentration compared to model 2 being multiplied by a factor almost 50 at a residence time of $18.5\times10^{-3}$s. Nevertheless, while we have chosen the reaction of this type with the largest flow rate, as the formation of five-membered rings is favored as discussed further in the text, the concentration of acetic acid is still 100 and 10000 times lower than those of formic acid and formaldehyde, respectively.

- Despite that the addition of OH• radicals to linear $C_3$-$C_5$ ketones has been found negligible under the conditions used by Tranter and Walker (2001), model 4 has been made in order to test the potential influence of the additions (204) and (205) of OH radicals to 2- and 3-heptanones, which are important primary products of the oxidation of n-heptane, as shown in fig. 8b:

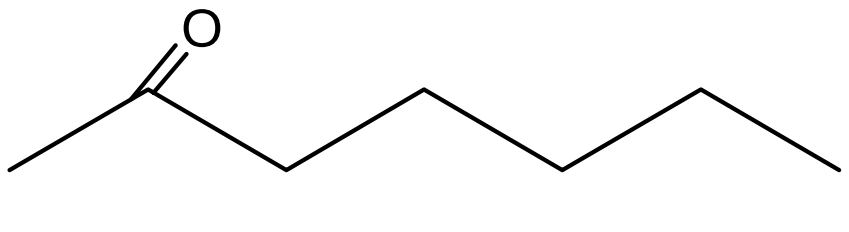

$+ OH\bullet \longrightarrow CH_3COOH + \bullet CH_3 + 2C_2H_4$ (204)

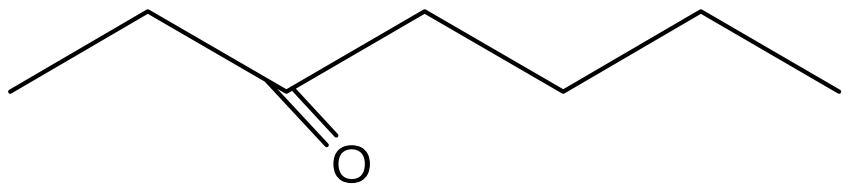

$+ OH\bullet \longrightarrow C_2H_5COOH + \bullet C_2H_5 + C_2H_4$ (205)

with rate parameters deriving from those proposed for reaction (198). In order to have a manageable size, the secondary mechanisms generated by the EXGAS software (Warth et al., 1998) involves lumped reactants (the molecules formed in the primary mechanism, with the same molecular formula and the same functional groups, are lumped into one unique species, without distinguishing between the different isomers) and includes global reactions. The different isomers of heptanone are then not distinguished, but both reactions (204) and (205) are considered, with rate constants reflecting the ratio between the flow rates of formation of 2- and 3-heptanones. The inclusion of these reactions has a spectacular effect on the production of acetic and propionic acids, which are now the major acids obtained, but with a concentration still about 400 times less than that of formaldehyde at a residence time of $18.5 \times 10^{-3}$s.

- In model 5, the decompositions of the ketohydroperoxides obtained by secondary reactions of cyclic ethers obtained in the oxidation of n-heptane (206) and iso-octane (207) have been replaced by the decompositions (208), (209) and (210), respectively:

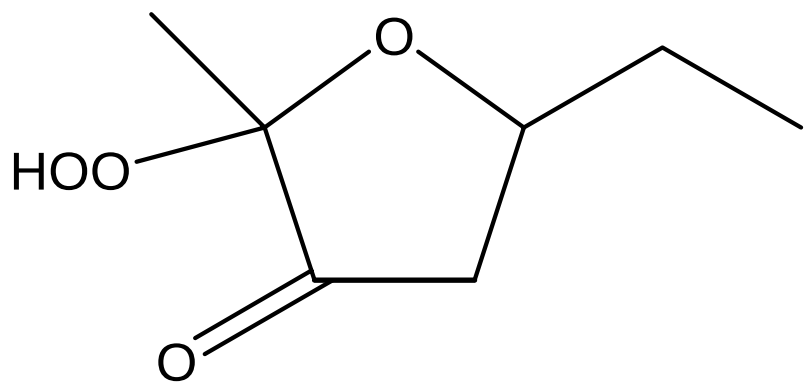

or

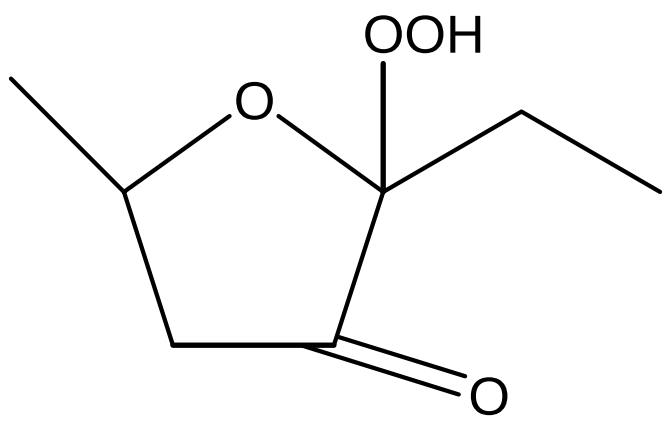

$\longrightarrow$ $OH\bullet + CO_2 + \bullet CH_2CHO + 2C_2H_4$ (206)

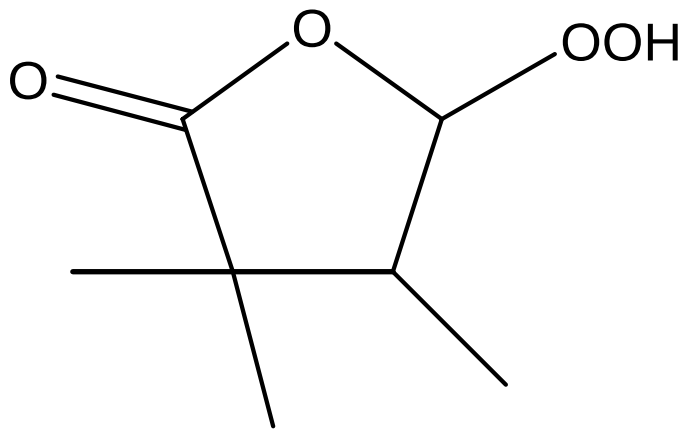

$\longrightarrow$ $OH\bullet + CO_2 + \bullet CHO + 3C_2H_4$ (207)

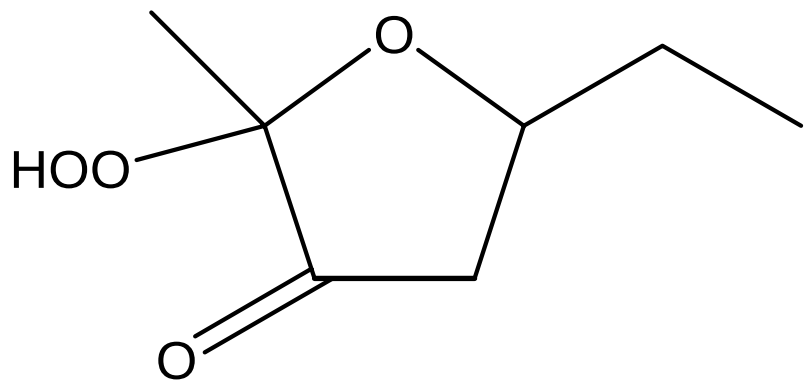

$\longrightarrow$ $OH\bullet + CO + CH_3COOH + \bullet C_2H_3 + C_2H_4$ (208)

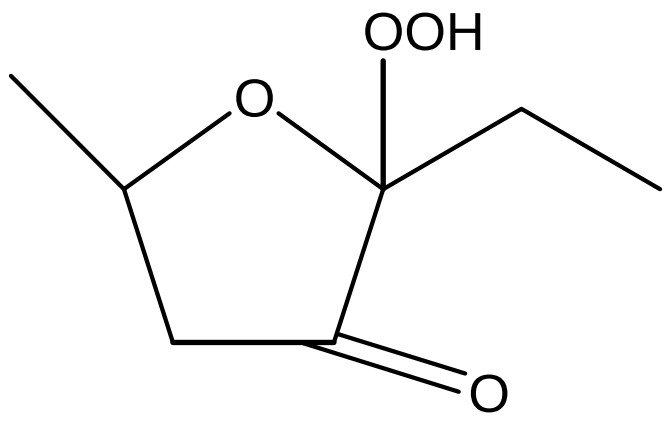

$\longrightarrow$ $OH\bullet + CO + C_2H_5COOH + \bullet C_3H_5$ (209)

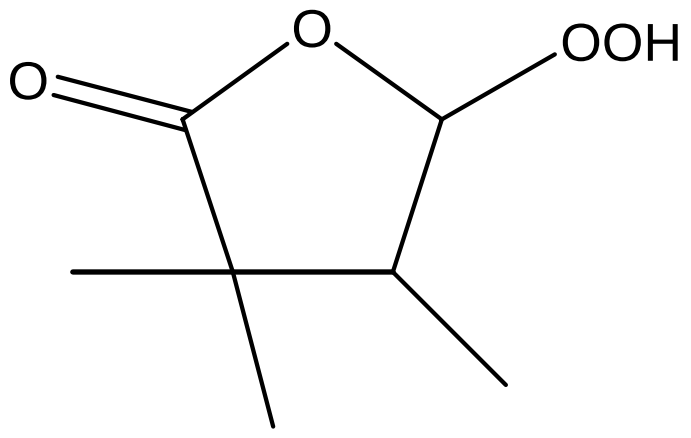

$\rightarrow$ $OH\bullet + CO + CH_3COOH + \bullet C_3H_5 + C_2H_4$ (210)

The decomposition of ketohydroperoxyde deriving from dimethyl ether has been proposed by Curran et al. (2000) to explain the formation of formic acid. The used rate parameters are an A-factor of $7.5x10^{15}$ $s^{-1}$, for reactions (208) and (209), and of $1.5x10^{16}$ $s^{-1}$ for reaction (210) and an activation energy of 42 kcal/mol, as proposed by Buda et al. (2005). Cyclic ethers react first by an H-abstraction and, as proposed by Glaude et al. (2000), the radicals obtained from cyclic ethers including more than 3 atoms can also react with oxygen. These reactions with oxygen involves the classical sequence of oxygen addition, isomerization, second oxygen addition, second isomerization and beta-scission to lead to the formation of ketohydroperoxydes, degenerate branching agents, which decomposes by reactions such as decompositions (206) and (207). As shown in fig. 8b, for a given alkane, the main produced cyclic ethers are five-membered rings (furans); the formation of four-membered (oxirans) and six-membered (pyrans) rings being less important. It is why we have tested the potential formation of acids only for the decompositions of the ketohydroperoxydes deriving from five-membered cyclic ethers. The changes in these reactions leads again to an important increase of the formation of acids, but the concentration of formadehyde still keeps a value about 80 times larger than that of acetic acid, the major acid obtained, at a residence time of $18.5 \times 10^{-3}$s.

**FIGURE 10**

The addition of OH• radicals to long chain ketones and the decomposition of the ketohydroperoxydes deriving from cyclic ethers seem to be the two most interesting ways to explain the presence of acids in exhaust gases. While the flow rate of formation of heptanones is larger than that of cyclic ethers, the production of acids is larger through this second type of compound, because ketones react mainly by H-abstraction under the studied conditions. While ways of formation of acids have been envisaged from the major oxygenated products of the oxidation of n-heptane and iso-octane, none of the tested models allows us to predict a higher formation of acids compared to formaldehyde. This could be due to a reaction channel which has not been envisaged and to the uncertainties linked with the single-zone modelling, but one could also consider the possibility of reactions in the exhaust line, may be also of heterogeneous reactions with the metallic parts.

## Conclusion

A new model has been proposed to simulate the formation of monocarboxylic acids from $C_1$ to $C_3$ in laminar premixed flames of propane. This model has allowed us to correctly reproduce the proportions between $C_1$, $C_2$ and $C_3$ acids experimentally measured. Under these conditions, $C_1$-$C_3$ acids are mainly derived from the aldehyde having the same structure, by addition of OH• radicals for formic acid and by combination of carbonyl and OH• radicals for heavier acids. The addition of OH• radicals to cyclopentadienone is a possible way to explain an enhancement of the amount of $C_3$ acid when toluene is added to the flame.

Under the conditions of the pre-ignition zone of an HCCI engine, the addition of OH• radicals to long chain ketones and the decomposition of the ketohydroperoxydes deriving from cyclic ethers could account for the formation of acetic and propionic acids as the major acids, but the obtained amounts are much lower than that of formaldehyde. These reactions should be considered as first tracks, which need to be more thoroughly investigated by means of additional experimental and theoretical work. A better understanding of the way of formation of monocarboxylic acids in the combustion chamber, but also in the exhaust line, would help to know whether internal combustion engines are really an important primary source of atmospheric emissions of these compounds.

**Acknowledgements**

C. Mounaïm-Rousselle and A. Dubreuil from the LME in Orléans are gratefully thanked for providing data for the simulations in an HCCI engine.

**References**

Aceves, S.M., Flowers, D.L., Espinosa-Loza , F., Martinez-Frias, J., Dec, J.E., Sjöberg, M., Dibble, R.W. and Hessel, R.P. (2004) Spatial analysis of emissions sources for HCCI combustion at low loads using a multi-zone model. SAE Paper 2004-01-1910.

Allara, D.L. and Shaw, R.J. (1980) A compilation of kinetic parameters for the thermal degradation of n-alkanes molecules**.** *J. Phys. Chem. Ref. Data,* **9**, 523.

Allendorf, M.D. (2006) HiTempThermo database. http://www.ca.sandia.gov/HiTempThermo/index.html.

Alzueta, M.U., Glarborg, P. and Dam-Johansen, K. (2000) Experimental and kinetic modeling study

of the oxidation of benzene. *Int. J. Chem. Kin.,* **32,** 498.

Atkinson, R., Baulch, D.L., Cox, R.A., Hampson, R.F., Kerr, J.A., Rossi, M.J. and Troe, J. (1997) Evaluated kinetic, photochemical and heterogeneous data for atmospheric chemistry: Supplement V - IUPAC Subcommittee on Gas Kinetic Data Evaluation for Atmospheric Chemistry. *J. Phys. Chem. Ref. Data* **26**, 521.

Atkinson, R., Baulch, D.L., Cox, R.A., Crowley, J.N., Hampson, R.F., Kerr, J.A., Hynes, R.G., Jenkin M.E., Rossi, M.J. and Troe, J. (2005) Evaluated kinetic and photochemical data for atmospheric chemistry: volume II – reactions of organic species. *Atmos. Chem. Phys. Discuss.,* **5**, 6295.

Barbé, P., Battin-Leclerc, F. and Côme, G.M. (1995) Experimental and modelling study of methane and ethane oxidation between 773 and 1573 K. *J. Chim. Phys.,* **92**, 1666.

Baulch, D.L., Bowman, C.T., Cobos, C.J., Cox, R.A., Just, Th., Kerr, J.A., Pilling, M.J., Stocker, D., Troe J., Tsang W., Walker, R.W. and Warnatz, J. (2005) Evaluated kinetic data for combustion modeling: supplement II. *J. Phys. Chem. Ref. Data*, **34**, 757.

Benson, S.W. (1076) *Thermochemical kinetics*, *2nd ed*. John Wiley, New York.

Biet, J., Delfau, J.L., Seydi, A. and Vovelle, C. (2005) Experimental and modeling study of lean premixed atmospheric-pressure propane/$O_2$/$N_2$ flames. *Combust. Flame,* **142**,197.

Borisov, A.A., Zamanskii, V.M., Konnov, A.A., and Skachkov G.I. (1990) Mechanism of high-temperature acetaldehyde oxidation. *Sov. J. Chem. Phys.*, **6**, 748.

Bounaceur, R., Da Costa, I., Fournet, R., Billaud, F. and Battin-Leclerc, F. (2005) Experimental and modeling study of the oxidation of toluene. *Int. J. Chem. Kin.,* **37**, 25.

Buda, F., Bounaceur, R., Warth, V., Glaude, P.A., Fournet, R. and Battin-Leclerc, F. (2005) Progress towards an unified detailed kinetic model for the autoignition of alkanes from $C_4$ to $C_{10}$ between 600 and 1200 K. *Combust. Flame,* **142**,170.

http://www.ensic.inpl-nancy.fr/DCPR/Anglais/GCR/exgas_distribution.htm

Burcat, A. and Ruscic B. (2006) Ideal Gas Thermochemical Database with updates from Active

Thermochemical Tables http://ftp.technion.ac.il/pub/supported/aetdd/thermodynamics.

Burrows, J.P., Moortgat, G.K., Tyndall, G.S., Cox, R.A., Jenkin, M.E., Hayman, G.D. and Veyret, B. (1989) Kinetics and mechanism of the photooxidation of formaldehyde. 2. Molecular modulation studies. *J. Phys. Chem.*, **93**, 2375

Calvert, G. and Stockwell, W.R. (1983) Acid generation in the troposphere by gas-phase chemistry. *Environ. Sci. Technol.*, **17**, 428A.

Chebbi, A. and Carlier, P. (1996) Carboxylic acids in the troposphere, occurrence, sources and sinks: a review. *Atmospheric Environment*, **30**, 4233.

Curran, H.J., Fisher, S.L. and Dryer F.L. (2000) The reaction kinetics of dimethyl ether. II: low-temperature oxidation in flow reactors. *Int. J. Chem. Kin.*, **32**, 741.

Da Costa, I., Fournet, R., Billaud, F. and Battin-Leclerc, F. (2003) Experimental and modeling study of the oxidation of benzene. *Int. J. Chem. Kin.,* **35**, 503.

Dagaut, P., Luche, J. and Cathonnet, M. (2000) Experimental and kinetic modeling of the reduction of NO by propene at 1 atm. *Combust. Flame*, **121**, 651.

DeMore, W.B., Sander, S.P., Golden, D.M., Hampson, R.F., Kurylo, M.J., Howard, C.J., Ravishankara, A.R., Kolb and C.E., Molina, M.J. (1997) Chemical kinetics and photochemical data for use in stratospheric modeling. Evaluation number 12 JPL Publication 97-4

Duan, X. and Page, M. (1995) Theoretical investigation of competing mechanisms in the thermal unimolecular decomposition of acetic acid and the hydration reaction of ketene. *J. Am. Chem. Soc.*, **117**, 5114.

Dubreuil, A., Foucher, F., Mounaïm-Rousselle, C., Dayma, G. and Dagaut, P. (2007) HCCI combustion: effect of NO in EGR. *Proc Combust. Inst.,* **31**, 2879.

Fournet, R., Baugé, J.C. and Battin-Leclerc, F. (1999) Experimental and modelling of oxidation of acetylene, propyne, allene and 1,3-butadiene *Int. J. Chem. Kin.,* **31**, 361.

Glaude, P.A., Battin-Leclerc, F., Fournet, R., Warth, V., Côme, G.M. and Scacchi, G. (2000) Construction and simplification of a model of the oxidation of alkanes. *Combust. Flame*, **122**, 451.

Gueniche, H.A., Glaude, P.A., Dayma, G., Fournet, R. and Battin-Leclerc, F. (2006) Rich methane premixed laminar flames doped by light unsaturated hydrocarbons – part 1: allene and propyne. *Combust. Flame,* **146**, 620.

Humpfer, R., Oser, H. and Grotheer, H-H. (1995) Formation of HCOH + $H_2$ through the reaction $CH_3$ + OH. Experimental evidence for a hitherto undetected product channel. *Int. J. Chem. Kinet.*, **27**, 577.

Hsu, D. S. Y., Shaub, W. M., Blackburn, M. and Lin, M. C. (1982) Thermal decomposition of formic acid at high temperatures in shock waves. *Proc. Combust. Inst.*, **19**, 89

Kee, R.J., Rupley, F.M. and Miller, J.A. (1993) Chemkin II. A fortran chemical kinetics package for the analysis of a gas-phase chemical kinetics. Sandia Laboratories Report, S 89-8009B.

Konnov, A.A. (2000) Detailed reaction mechanism for small hydrocarbons combustion. Release 0.5. 2000. http://homepages.vub.ac.be/~akonnov.

Konnov, A.A. (2006) Towards a unified C1-C3 kinetic mechanism for combustion modelling over extended temperature and pressure range. Proceed. of the 19th International Symposium on Gas Kinetics (GK 2006) Orléans, France July 22-27 pp. 259-260.

Larson, C.W., Stewart, P.H. and Golden, D.M. (1988) Pressure and temperature dependence of reactions proceeding via a bound complex. An approach for combustion and atmospheric chemistry modelers. Application to HO + CO = [HOCO] = H + $CO_2$. *Int. J. Chem. Kinet.*, **20**, 27.

Legrand, M., Preunkert, S., Wagenbach, D., Cachier, H. and Puxbaum H., J. (2003) A historical record of formate and acetate from a high-elevation Alpine glacier: Implication for their natural versus anthropogenic budgets at the European Scale. *J. Geophys. Res.,* **108**, 4788.

Marinov, N.M., Pitz, W.J., Westbrook, C.K., Castaldi, M.J. and Senkan S.M. (1996) Modeling aromatic and polycyclic aromatic hydrocarbon formation in premixed methane and ethane flames. *Combust. Sci. Technol.*, **116-117**, 211

Miller, J.A. and Melius, C.F. (1989) A theoretical analysis of the reaction between hydroxyl and acetylene. *Proc. Combust. Inst.*, **22**, 1031.

Moortgat, G., Veyret, B. and Lesclaux, R. ((1989) Absorption spectrum and kinetics of reactions of the acetylperoxy radical. *J. Phys. Chem.*, **93**, 2362

Muller, C., Michel, V., Scacchi, G. and Côme, G.M. (1995) A computer program for the evaluation of thermochemical data of molecules and free radicals in the gas phase. *J. Chim. Phys.*, **92**, 1154.

Orlando, J.J. and Tyndall, G.S. (2001) The Atmospheric Chemistry of the HC(O)CO Radical. *Int. J. Chem. Kin.*, **33**,149.

Petty, J.T., Harrison, J.A. and Moore, C.B. (1993) Reactions of *trans*-HOCO studied by infrared spectroscopy. *J. Phys. Chem.*, **97**, 11194.

Saito, K., Kakumoto, T., Kuroda, H., Torii, S. and Imamura, A. (1984) Thermal unimolecular decomposition of formic acid. *J. Chem. Phys.*, **80**, 4989.

Senosiain, J.P., Musgrave, C.B. and Golden, D.M. Temperature and pressure dependence of the reaction of OH and CO: Master equation modeling on a high-level potential energy surface. (2003) *Int. J. Chem. Kin.*, **35**, 464.

Sivakumaran, V., Hölscher, D., Dillon, T.J., and Crowley, J.N. (2003) Reaction between OH and HCHO: temperature dependent rate coefficients (202–399 K) and product pathways (298 K). *Phys. Chem. Chem. Phys.,* **5**, 4821.

Stoeckel, F., Schuh, M.D., Goldstein, N. and Atkinson, G.H. (1985) Time-resolved intracavity laser spectroscopy: 266 nm photodissociation of acetaldehyde vapor to form HCO. *Chem. Phys.*, **95**, 135

Talukdar R.J., Gierczak, T., McCabe, D.C. and Ravishankara, A.R. (2003). Reaction of hydroxyl radical with acetone. 2. Products and reaction mechanism. *J. Phys. Chem. A,* **107**, 5021.

Taylor, P.H., Yamada, T. and Marshall, P. (2006) The reaction of OH with acetaldehyde and deuterated acetaldehyde: further insight into the reaction mechanism at both low and elevated temperatures. *Int. J. Chem. Kin.,* **38**, 489.

Thomson, M.J., Lucas, D., Koshland, C.P., Sawyer, R.F., Wu, Y.P and Bozzelli, J.W. (1994) An experimental and numerical study of the high-temperature oxidation of 1,1,1-$C_2H_3Cl_3$. *Combust. Flame*, **98**,155.

Tomas, A., Villenave, E. and Lesclaux, R. (2001) Reactions of the $HO_2$ radicals with $CH_3CHO$ and $CH_3C(O)O_2$ in the gas phase. *J. Phys. Chem. A,* **105,** 3505.

Touchard, S., Fournet, R., Glaude, P.A., Warth, V., Battin-Leclerc, F., Vanhove, G., Ribaucour, M. and Minetti R. (2005) Modeling of the oxidation of large alkenes at low temperature. *Proc. Combust. Inst.,* **30**, 1073.

Tranter, R.S. and Walker, R.W. (2001) Rate constant for H and OH attack on propanone, butanone and petan-3-one at 753 K, and the oxidation chemistry of the radicals formed. *Phys. Chem. Chem. Phys.,* **3**, 1262.

Troe, J. (1974) Fall-off curves of unimolecular reaction. *Ber. Buns. Phys. Chem.,* **78**, 478.

Tsang, W. and Hampson, R.F. (1986) Chemical kinetic data base for combustion chemistry. Part I. Methane and related compounds. *J. Phys. Chem. Ref. Data*, **15**, 1087.

Tyndall, G.S., Orlando, J.J., Wallington, T.J., Hurley, M.D., Goto, M. and Kawasaki M. (2002). Mechanism of the reaction of OH radicals with acetone and acetaldehyde at 251 and 296 K. *Phys. Chem. Chem. Phys.,* **4**, 2189.

Vandenberk,S., Vereecken, L. and Peeters, J. (2002). The acetic acid forming channel in the acetone + OH reaction: A combined experimental and theoretical investigation. *Phys. Chem. Chem.. Phys.,* **4**, 461.

Warth, V., Stef, N., Glaude, P.A., Battin-Leclerc, F., Scacchi, G. and Côme, G.M. (1998) Computed aided design of gas-phase oxidation mechanisms: Application to the modelling of normal-butane oxidation. *Combust. Flame,* **114**, 81.

Wallington,T.J., Dagaut,P. and Kurylo,M.J. (1992) Ultraviolet absorption cross sections and reaction kinetics and mechanisms for peroxy radicals in the gas phase. *Chem. Rev.*, **92**, 667.

Watkins, K. W. and Thompson, W. W. (1973) Addition of ethyl radicals to carbon monoxide. Kinetic and thermochemical properties of the propionyl radical. *Int. J. Chem. Kin.*, **5**, 791.

Wilk, R.D., Cernansky, N.P., Pitz, W.J., and Westbrook, C.K. (1989) Propene oxidation at low and intermediate temperatures: A detailed chemical kinetic study. *Combust. Flame*, **77**, 145.

Zervas, E., Montagne, X. and Lahaye, J. (1999) Collection and analysis of organic acids in exhaust gas. Comparison of different methods. *Atmospheric Environment,* **33**, 4953.

Zervas, E., Montagne, X. and Lahaye, J. (2001a) Emission of specific pollutants from a compression ignition engine. Influence of fuel hydrotreatment and fuel/air equivalence ratio. *Atmospheric Environment,* **35**, 1301.

Zervas, E., Montagne, X. and Lahaye, J. $C_1$-$C_5$ Organic acid emissions from an SI engine: Influence of fuel and air/fuel equivalence ratio. (2001b) *Environ. Sci. Technol.,* **35**, 2746.

Zervas, E., Montagne, X. and Lahaye, J. (2002) Emission of alcohols and carbonyl compounds from a spark ignition engine. Influence of fuel and air/fuel equivalence ratio. *Environ. Sci. Technol.,* **36**, 2414.

Zervas, E., Montagne, X. and Lahaye, J. (2003) Emissions of regulated pollutants from a spark ignition engine. Influence of fuel and air/fuel equivalence ratio. *Environ. Sci. Technol.,* **37**, 3232.

Zervas, E. (2005) Formation of organic acids from propane, isooctane and toluene/isooctane flames. *Fuel,* **84**, 691.

**TABLE I: NAMES, FORMULAE AND HEATS OF FORMATION (298K), ENTROPY (298K) AND SPECIFIC HEAT (300K) FOR SPECIES INVOLVED IN TABLE II, WHICH WERE NOT CONSIDERED BY GUENICHE ET AL. (2006).**
**Thermochemical data are given in cal, mol, K units.**

| Name | Formula | $\Delta H_f$ /1000 | S | $C_p$ | Name | Formula | $\Delta H_f$ /1000 | S | $C_p$ |
|---|---|---|---|---|---|---|---|---|---|
| | | | | | **$C_1$ species** | | | | |
| HCOOH | H-C(=O)-OH | -90.5 | 59.1 | 9.9 | HCOH[a] | H-C(-OH) •• | 49.5 | 57.2 | 10.5 |
| HOCO | HO-C(=O) • | -43.3 | 60.2 | 10.4 | HCOO | H-C(=O)-O• | -36.0 | 57.3 | 9.6 |
| HOCHOH | HO-CH(-OH) • | -48.1 | 66.7 | 11.0 | HOCH2O | HO-CH2-O• | -39.1 | 63.2 | 12.5 |
| CH2(OH)2 | HO-CH2-OH | -91.0 | 64.5 | 11.3 | HOCH2OO | HO-CH2-O-O• | -37.1 | 69.1 | 16.0 |
| HOCH2OOH | HO-CH2-O-OH | -75.2 | 74.1 | 15.9 | | | | | |
| | | | | | **$C_2$ species** | | | | |
| CH3CO2H | CH3-C(=O)-OH | -103.0 | 67.7 | 15.2 | HCCOH | H-C≡C-OH | 22.3 | 59.5 | 13.8 |
| C2H2OH | H-C(•)=C-OH | 30.3 | 62.4 | 13.9 | H2CCOH | H2C=C (-OH)• | 28.2 | 64.2 | 13.3 |
| CH2CHOH | H2C=C-H(-OH) | -29.8 | 69.3 | 14.6 | CHOCO | O=CH-C=O | -13.8 | 39.1 | 11.6 |
| CHOCHO | O=CH-CH=O | -50.7 | 65.1 | 14.3 | HOCH2CO | HO-CH2-C(=O)• | -39.5 | 75.6 | 16.0 |
| HOCH2CHO | HO-CH2-C-H(=O) | -74.8 | 74.1 | 17.0 | HOCHCHO | HO-CH(•)-CH=O | -32.5 | 72.1 | 16.7 |
| CHCHOOH | HC(•)=CH-O-OH | 51.6 | 74.1 | 18.2 | CH3CO2 | CH3-C(=O)-O• | -50.3 | 65.6 | 14.0 |
| C2H3O2 | CH2=CH(-O-O•) | 27.9 | 68.5 | 17.1 | C2H3OOH | CH2=CH(-O-OH) | -72.5 | 73.0 | 18.8 |
| CH3CHOOH | CH3-CH(•)-O-OH | 74.5 | 74.5 | 20.2 | | | | | |
| | | | | | **$C_3$ species** | | | | |
| C2H5CO2H | CH3-CH2-C(=O)-OH | -106.0 | 76.9 | 21.9 | C2H3CO2H | CH2=CH-C(=O)-OH | -80.4 | 75.2 | 18.7 |
| C2H5CO | CH3-CH2-C(=O) • | -86.1 | 73.1 | 18.3 | C2H5CO2 | CH3-CH2-C(=O)-O• | -52.0 | 75.0 | 19.9 |
| C2H5COOO | CH3-CH2-C(=O)-O-O• | -51.4 | 85.6 | 21.8 | C2H5COOOH | CH3-CH2-C(=O)-O-O-H | -86.5 | 90.1 | 23.5 |
| C2H3CO2 | CH2=CH-C(=O)-O• | -26.4 | 73.3 | 16.7 | | | | | |

[a] : Triplet state.

**TABLE II: COMPLETE LIST OF THE REACTIONS CONCERNING $C_1$, $C_2$ AND $C_3$ OXYGENATED COMPOUNDS WHICH WERE NOT PART OF THE MECHANISM FOR THE OXIDATION OF ALLENE AND PROPYNE PROPOSED BY GUENICHE ET AL. (2006). THESE REACTIONS WERE MAINLY PROPOSED BY KONNOV (2006).**

**The rate constants are given ($k=A\ T^n \exp(-E_a/RT)$) in cc, mol, s, cal units.**

| No | Reaction | A | n | Ea | Reference |
|---|---|---|---|---|---|
| | | **$C_1$ species** | | | |
| 1 | CH3+OH=HCOH+H2 | 2.28E+10 | -0.12 | -415 | Baulch et al., 2005 |
| 2 | HCOH+OH=CO2+H2+H | 1.08E+13 | 0 | 0 | Humpfer et al., 1995 |
| 3 | HCOH+HCHO=HOCH2CHO | 1.00E+11 | 0 | 0 | Estimate |
| 4 | HCOH+CH3=CH2CHOH+H | 7.00E+13 | 0 | 0 | Estimate |
| 5 | C2H4+HCOH=C2H6+CO | 1.00E+11 | 0 | 0 | Estimate |
| 6 | CO+OH(+M)=HOCO(+M) | 1.20E+07 | 1.83 | -236 | Senosiain et al., |
| | LOW | 7.20E+25 | -3.85 | 1550 | (2003) |
| | TROE /0.6 10. 100000./ | | | | |
| 7 | HOCO(+M)=H+CO2(+M) | 1.74E+12 | 0.307 | 32930 | Larson et al., 1988 |
| | LOW | 2.29E+26 | -3.02 | 35070 | |
| 8 | HOCO+O2=HO2+CO2 | 1.26E+12 | 0 | 0 | Petty et al., 1993 |
| 9 | CH3CO+OH=CH3+HOCO | 3.00E+13 | 0 | 0 | Tsang and Hampson, 1986 |
| 10 | HCOO+M=H+CO2+M | 8.70E+15 | 0 | 14400 | Same rate constant as for the decomposition of CH3CO2 |
| 11 | HCOO+O2=CO2+HO2 | 1.00E+11 | 0 | 0 | Estimate |
| 12 | HCOOH+M=CO+H2O+M | 2.10E+14 | 0 | 40400 | Saito et al., 1984 |
| 13 | HCOOH+M=CO2+H2+M | 1.50E+16 | 0 | 57000 | Hsu et al., 1982 |
| 14 | HCOOH+OH=HOCO+H2O | 1.00E+11 | 0 | 0 | Estimate |
| 15 | HCOOH+OH=HCOO+H2O | 1.00E+11 | 0 | 0 | Estimate |
| 16 | HCOOH+HCOO=HOCO+HCOOH | 1.00E+11 | 0 | 0 | Estimate |
| 17 | HCHO+OH=HOCH2O | 3.40E+06 | 1.18 | -400 | See present work |
| 18 | HOCH2O=HCOOH+H | 1.00E+14 | 0 | 14900 | Estimate |
| 19 | HOCH2O+O2=HO2+HCOOH | 2.10E+10 | 0 | 0 | Estimate |
| 20 | HOCH2O+HCHO=CH2(OH)2+CHO | 6.00E+08 | 0 | 0 | Estimate |
| 21 | CH2(OH)2(+M)=CH2OH+OH(+M) | 1.70E+16 | 0 | 90885 | Same reactivity |
| | LOW | 6.60E+16 | 0 | 65730 | assumed as CH3OH |
| | TROE /0.82 200.0 1438.0/ | | | | (Konnov, 2000) |
| 22 | CH2(OH)2(+M)=HOCHOH+H(+M) | 1.38E+16 | 0 | 95950 | Same reactivity |
| | LOW | 5.35E+16 | 0 | 70800 | assumed as CH3OH |
| | TROE /0.82 200.0 1438.0/ | | | | |
| 23 | CH2(OH)2+H=HOCHOH+H2 | 1.64E+07 | 2.0 | 4520 | Same reactivity assumed as CH3OH |
| 24 | CH2(OH)2+H=H2+HOCH2O | 4.00E+13 | 0 | 6095 | Same reactivity assumed as CH3OH |
| 25 | CH2(OH)2+O=HOCHOH+OH | 1.63E+13 | 0 | 5030 | Same reactivity assumed as CH3OH |
| 26 | CH2(OH)2+O=HOCH2O+OH | 1.00E+13 | 0 | 4680 | Same reactivity assumed as CH3OH |
| 27 | CH2(OH)2+OH=HOCHOH+H2O | 1.44E+06 | 2.0 | -840 | Same reactivity assumed as CH3OH |
| 28 | CH2(OH)2+OH=HOCH2O+H2O | 1.00E+13 | 0 | 1700 | Same reactivity assumed as CH3OH |
| 29 | CH2(OH)2+O2=HOCHOH+HO2 | 2.05E+13 | 0 | 44900 | Same reactivity assumed as CH3OH |
| 30 | CH2(OH)2+HO2=HOCHOH+H2O2 | 9.64E+10 | 0 | 12580 | Same reactivity assumed as CH3OH |
| 31 | CH2(OH)2+B5CH2=CH3+HOCHOH | 3.19E+01 | 3.2 | 7200 | Same reactivity assumed as CH3OH |
| 32 | CH2(OH)2+B5CH2=CH3+HOCH2O | 1.44E+01 | 3.1 | 6900 | Same reactivity assumed as CH3OH |
| 33 | CH2(OH)2+CH3=HOCHOH+CH4 | 3.19E+01 | 3.17 | 7172 | Same reactivity assumed as CH3OH |
| 34 | CH2(OH)2+CH3=HOCH2O+CH4 | 1.45E+01 | 3.1 | 6935 | Same reactivity assumed as CH3OH |
| 35 | HCHO+HO2=HOCH2OO | 5.80E+09 | 0 | -1242 | Atkinson et al., 1997 |
| 36 | CH2OH+O2=HOCH2OO | 1.50E+14 | -1.0 | 0 | Rate constant taken as 10% as that used by Konnov (2000) for the channel leading to HCHO at low temperature |
| 37 | HOCH2OO+HO2=HCOOH+H2O+O2 | 3.37E+09 | 0 | -4570 | Atkinson et al. 1997 |
| 38 | HOCH2OO+HOCH2OO= HCOOH+CH2(OH)2+O2 | 3.40E+10 | 0 | -1490 | Wallington et al. 1992 |
| 39 | HOCH2OO+HOCH2OO= HOCH2O+HOCH2O+O2 | 3.31E+12 | 0 | 0 | Burrows et al. 1989 |
| 40 | HOCH2OO+H=HOCH2O+OH | 9.60E+13 | 0 | 0 | Same reactivity assumed as CH3O2 (Konnov, 2000) |
| 41 | HOCH2OO+O=HOCH2O+O2 | 3.60E+13 | 0 | 0 | Same reactivity assumed as CH3O2 |
| 42 | HOCH2OO+OH=CH2(OH)2+O2 | 6.00E+13 | 0 | 0 | Same reactivity assumed as CH3O2 |
| 43 | HOCH2OO+OH=HOCH2O+HO2 | 3.00E+12 | 0 | 0 | Same reactivity assumed as CH3O2 |
| 44 | HOCH2OO+HO2=HOCH2OOH+O2 | 2.52E+11 | 0 | -1490 | Same reactivity assumed as CH3O2 |
| 45 | HOCH2OO+HO2=CH2OO+H2O+O2 | 9.60E+08 | 0 | -3440 | Same reactivity assumed as CH3O2 |
| 46 | HOCH2OO+H2O2=HOCH2OOH+HO2 | 2.40E+12 | 0 | 9940 | Same reactivity assumed as CH3O2 |
| 47 | HOCH2OO+B5CH2=HCHO+HOCH2O | 1.80E+13 | 0 | 0 | Same reactivity assumed as CH3O2 |
| 48 | HOCH2OO+CH3=CH3O+HOCH2O | 5.00E+12 | 0 | -1410 | Same reactivity assumed as CH3O2 |
| 49 | CH4+HOCH2OO=CH3+HOCH2OOH | 1.81E+11 | 0 | 18480 | Same reactivity assumed as CH3O2 |
| 50 | HOCH2OO+CO=HOCH2O+CO2 | 1.00E+14 | 0 | 24000 | Same reactivity assumed as CH3O2 |
| 51 | HCHO+HOCH2OO=CHO+HOCH2OOH | 2.00E+12 | 0 | 11660 | Same reactivity assumed as CH3O2 |
| 52 | HOCH2OO+CH3O=HCHO+HOCH2OOH | 3.00E+11 | 0 | 0 | Same reactivity assumed as CH3O2 |
| 53 | CH3OH+HOCH2OO=CH2OH+HOCH2OOH | 1.81E+12 | 0 | 13700 | Same reactivity assumed as CH3O2 |
| 54 | HOCH2OO+CH3OH=HOCH2OOH+CH3O | 2.80E+11 | 0 | 18800 | Same reactivity assumed as CH3O2 |
| 55 | HOCH2OOH=HOCH2O+OH | 6.00E+14 | 0 | 42300 | Same reactivity assumed as CH3O2H (Konnov 2000) |
| 56 | HOCH2OOH+H=H2+HOCH2OO | 8.79E+10 | 0 | 1860 | Same reactivity assumed as CH3O2H |
| 57 | HOCH2OOH+H=H2O+HOCH2O | 7.27E+10 | 0 | 1860 | Same reactivity assumed as CH3O2H |
| 58 | HOCH2OOH+O=OH+HOCH2OO | 2.80E+13 | 0 | 6400 | Same reactivity assumed as CH3O2H |
| 59 | HOCH2OOH+OH=HOCH2OO+H2O | 1.08E+12 | 0 | -437 | Same reactivity assumed as CH3O2H |
| | **$C_2$ species** | | | | |
| 60 | C2H2+OH=H+HCCOH | 5.06E+05 | 2.3 | 13500 | Miller and Melius, 1989 |
| 61 | HCCOH+H=H+CH2CO | 1.00E+13 | 0 | 0 | Thomson et al., 1994 |
| 62 | HCCOH+O=CHCO+OH | 1.00E+10 | 0 | 0 | Estimate |
| 63 | C2H2+OH(+M)=C2H2OH(+M) | 2.28E+13 | 0 | 1800 | Atkinson et al., |
| | LOW | 9.34E+21 | -1.5 | | 1997 |
| 64 | C2H2OH=CH2CO+H | 5.00E+15 | 0 | 28000 | Dagaut et al., 2000 |
| 65 | C2H2OH+H=CH2CO+H2 | 2.00E+13 | 0 | 4000 | Dagaut et al., 2000 |
| 66 | C2H2OH+O=CH2CO+OH | 2.00E+13 | 0 | 4000 | Dagaut et al., 2000 |
| 67 | C2H2OH+OH=CH2CO+H2O | 1.00E+13 | 0 | 2000 | Dagaut et al., 2000 |
| 68 | C2H2OH+O2=CHOCHO+OH | 1.00E+12 | 0 | 5000 | Dagaut et al., 2000 |
| 69 | C2H2OH+O2=CHO+CO2+H2 | 4.00E+12 | 0 | -250 | Dagaut et al., 2000 |
| 70 | H2CCOH=CH2CO+H | 5.00E+15 | 0 | 28000 | Same reactivity assumed as C2H2OH (Konnov, 2000) |
| 71 | H2CCOH+H=CH2CO+H2 | 2.00E+13 | 0 | 4000 | Same reactivity assumed as C2H2OH |
| 72 | H2CCOH+O=CH2CO+OH | 2.00E+13 | 0 | 4000 | Same reactivity assumed as C2H2OH |
| 73 | H2CCOH+OH=CH2CO+H2O | 1.00E+13 | 0 | 2000 | Same reactivity assumed as C2H2OH |
| 74 | H2CCOH+O2=CHOCHO+OH | 1.00E+12 | 0 | 5000 | Same reactivity assumed as C2H2OH |
| 75 | H2CCOH+O2=CHO+CO2+H2 | 4.00E+12 | 0 | -250 | Same reactivity assumed as C2H2OH |
| 76 | CH2CHOH+H=C2H2OH+H2 | 1.64E+07 | 2.0 | 10000 | Same reactivity assumed as CH3OH (Konnov, 2000) |
| 77 | CH2CHOH+H=H2CCOH+H2 | 4.00E+13 | 0 | 10000 | Same reactivity assumed as CH3OH |
| 78 | CH2CHOH+H=CH2CHO+H2 | 4.00E+13 | 0 | 10000 | Same reactivity assumed as CH3OH |
| 79 | CH2CHOH+O=C2H2OH+OH | 1.63E+13 | 0 | 12000 | Same reactivity assumed as CH3OH |
| 80 | CH2CHOH+O=H2CCOH+OH | 1.00E+13 | 0 | 10000 | Same reactivity assumed as CH3OH |
| 81 | CH2CHOH+O=CH2CHO+OH | 1.00E+13 | 0 | 5000 | Same reactivity assumed as CH3OH |
| 82 | CH2CHOH+OH=H2CCOH+H2O | 1.00E+13 | 0 | 1700 | Same reactivity assumed as CH3OH |
| 83 | CH2CHOH+OH=C2H2OH+H2O | 1.44E+06 | 2.0 | -840 | Same reactivity assumed as CH3OH |
| 84 | CH2CHOH+OH=CH2CHO+H2O | 1.00E+13 | 0 | 1700 | Same reactivity assumed as CH3OH |
| 85 | CH2CHOH+O2=H2CCOH+HO2 | 2.05E+13 | 0 | 64000 | Same reactivity assumed as CH3OH |
| 86 | CH2CHOH+HO2=H2CCOH+H2O2 | 9.64E+10 | 0 | 25000 | Same reactivity assumed as CH3OH |
| 87 | CH2CHOH+CH3=C2H2OH+CH4 | 3.19E+01 | 3.17 | 10000 | Same reactivity assumed as CH3OH |
| 88 | CH2CHOH+CH3=H2CCOH+CH4 | 1.45E+01 | 3.1 | 10000 | Same reactivity assumed as CH3OH |
| 89 | CH2CHOH+CH3=CH2CHO+CH4 | 1.45E+01 | 3.1 | 7000 | Same reactivity assumed as CH3OH |
| 90 | CHOCO=CHO+CO | 1.42E+12 | 0 | 6280 | Orlando and Tyndall, 2001 |
| 91 | CHOCO+O2=CO+CO+HO2 | 3.00E+12 | 0 | 0 | Rate constant taken as that used by Konnov (2000) for HCO+O2=CO+HO2 |
| 92 | CHO+CHO=CHOCHO | 3.00E+13 | 0 | 0 | Stoeckel et al., 1985 |
| 93 | C2H3+O2=CHOCHO+H | 1.47E+23 | -3.83 | 6240 | Thompson et al., 1994 |
| 94 | CH2CHO+O2=CHOCHO+OH | 2.76E+12 | 0 | 3000 | Dagaut et al., 2000 |
| 95 | CHOCHO=HCHO+CO | 1.17E+16 | -1.28 | 50937 | Dagaut et al., 2000 |
| 96 | CHOCHO=CO+CO+H2 | 6.52E+39 | -7.70 | 67469 | Dagaut et al., 2000 |
| 97 | CHOCHO+H=HCHO+CHO | 1.00E+12 | 0 | 0 | Dagaut et al., 2000 |
| 98 | CHOCHO+O=CHOCO+OH | 7.24E+12 | 0 | 1970 | Dagaut et al., 2000 |
| 99 | CHOCHO+OH=CHOCO+H2O | 6.60E+12 | 0 | 0 | Atkinson et al., 1997 |
| 100 | CHOCHO+O2=CHO+CO+HO2 | 6.30E+13 | 0 | 30000 | Dagaut et al., 2000 |
| 101 | CHOCHO+HO2=CHOCO+H2O2 | 1.70E+12 | 0 | 10700 | Dagaut et al., 2000 |
| 102 | CHOCHO+CH3=CHOCO+CH4 | 1.74E+12 | 0 | 8440 | Dagaut et al., 2000 |
| 103 | CH3CO2+M=CH3+CO2+M | 8.70E+15 | 0 | 14400 | Wilk et al. 1989 |
| 104 | CH2OH+CO+M=HOCH2CO+M | 1.00E+14 | 0 | 3000 | Same reactivity assumed as CH3CO (Konnov, 2000) |
| 105 | HOCH2CO+H=CH2OH+CHO | 2.15E+13 | 0 | 0 | Same reactivity assumed as CH3CO |
| 106 | HOCH2CO+O=CH2OH+CO2 | 1.58E+14 | 0 | 0 | Same reactivity assumed as CH3CO |
| 107 | HOCH2CO+OH=CH2OH+HOCO | 3.00E+13 | 0 | 0 | Same reactivity assumed as CH3CO |
| 108 | HOCH2CO+O2=HOCH2O+CO2 | 4.44E+10 | 0 | -1080 | Same reactivity assumed as CH3CO |
| 109 | HOCH2CO+HO2=CH2OH+CO2+OH | 3.00E+13 | 0 | 0 | Same reactivity assumed as CH3CO |
| 110 | HOCH2CO+CH3=C2H5OH+CO | 3.30E+13 | 0 | 0 | Same reactivity assumed as CH3CO |
| 111 | HOCH2CO+CHO=HOCH2CHO+CO | 9.00E+12 | 0 | 0 | Same reactivity assumed as CH3CO |
| 112 | HOCH2CO+HCHO=HOCH2CHO+CHO | 1.80E+11 | 0 | 12900 | Same reactivity assumed as CH3CO |
| 113 | HOCH2CO+CH3O=HCHO+HOCH2CHO | 6.00E+12 | 0 | 0 | Same reactivity assumed as CH3CO |
| 114 | HOCH2CO+CH3OH=HOCH2CHO+CH2OH | 4.85E+03 | 3.0 | 12340 | Same reactivity assumed as CH3CO |
| 115 | HOCHCHO=HOCH2CO | 1.00E+13 | 0 | 47000 | Same reactivity assumed as CH2CHO (Konnov, 2000) |
| 116 | HOCHCHO=CH2OH+CO | 7.80E+41 | -9.15 | 46900 | Same reactivity assumed as CH2CHO |
| 117 | HOCHCHO+O=HCOOH+CHO | 9.60E+06 | 1.83 | 220 | Same reactivity assumed as CH2CHO |
| 118 | HOCHCHO+HO2=HCOOH+CHO+OH | 1.10E+13 | 0 | 0 | Same reactivity assumed as CH2CHO |
| 119 | HOCHCHO+CH3=SC2H5O+CHO | 4.90E+14 | -0.5 | 0 | Same reactivity assumed as CH2CHO |
| 120 | CHCHOOH=CH2CO+OH | 3.16E+11 | 0 | 19500 | Same reactivity assumed as CH2CH2OOH (Konnov, 2000) |
| 121 | CHCHOOH=HOCHCHO | 2.60E+12 | 0 | 20000 | Same reactivity assumed as CH2CH2OOH |
| 122 | C2H3+O2=C2H3O2 | 6.00E+12 | 0 | 0 | Same reactivity assumed as C2H5O2 (Konnov, 2000) |
| 123 | C2H3O2=CHCHOOH | 2.00E+12 | 0 | 40000 | Same reactivity assumed as C2H5O2 |
| 124 | C2H3O2=C2H2+HO2 | 5.00E+11 | 0 | 40000 | Same reactivity assumed as C2H5O2 |
| 125 | C2H3O2+H2=C2H3OOH+H | 3.00E+12 | 0 | 21000 | Same reactivity assumed as C2H5O2 |
| 126 | C2H3O2+HO2=C2H3OOH+O2 | 3.00E+11 | 0 | -1000 | Same reactivity assumed as C2H5O2 |
| 127 | C2H3O2+HO2=CH2CHO+OH+O2 | 1.80E+12 | 0 | 0 | Same reactivity assumed as C2H5O2 |
| 128 | C2H3O2+H2O2=C2H3OOH+HO2 | 4.00E+11 | 0 | 11000 | Same reactivity assumed as C2H5O2 |
| 129 | C2H3O2+CH3=CH2CHO+CH3O | 3.00E+13 | 0 | 0 | Same reactivity assumed as C2H5O2 |
| 130 | CH4+C2H3O2=CH3+C2H3OOH | 1.00E+13 | 0 | 21000 | Same reactivity assumed as C2H5O2 |
| 131 | HCHO+C2H3O2=CHO+C2H3OOH | 1.00E+11 | 0 | 9000 | Same reactivity assumed as C2H5O2 |
| 132 | C2H3O2+CH3OH=C2H3OOH+CH3O | 1.00E+11 | 0 | 19000 | Same reactivity assumed as C2H5O2 |
| 133 | CH3OH+C2H3O2=CH2OH+C2H3OOH | 6.00E+12 | 0 | 20000 | Same reactivity assumed as C2H5O2 |
| 134 | C2H3O2+CH3OO= CH2CHO+CH3O+O2 | 1.00E+11 | 0 | 0 | Same reactivity assumed as C2H5O2 |
| 135 | C2H3O2+CH3OOH=C2H3OOH+CH3OO | 1.00E+12 | 0 | 16000 | Same reactivity assumed as C2H5O2 |
| 136 | C2H4+C2H3O2=C2H3+C2H3OOH | 1.00E+13 | 0 | 31000 | Same reactivity assumed as C2H5O2 |
| 137 | C2H3O2+C2H4=CH2CHO+C2H4O#3 | 2.00E+12 | 0 | 18000 | Same reactivity assumed as as C2H5O2 |
| 138 | C2H3O2+C2H6=C2H3OOH+C2H5 | 1.00E+13 | 0 | 21000 | Same reactivity assumed as C2H5O2 |
| 139 | C2H3O2+CH2CO=C2H3OOH+CHCO | 1.00E+12 | 0 | 25000 | Same reactivity assumed as C2H5O2 |
| 140 | CH3CHO+C2H3O2=CH3CO+C2H3OOH | 1.00E+11 | 0 | 10000 | Same reactivity assumed as C2H5O2 |
| 141 | C2H3O2+CH3CHO=C2H3OOH+CH2CHO | 1.00E+12 | 0 | 20000 | Same reactivity assumed as C2H5O2 |
| 142 | C2H3O2+C2H3O2= CH2CHO+CH2CHO+O2 | 4.00E+10 | 0 | 0 | Same reactivity assumed as C2H5O2 |
| 143 | C2H3OOH=OH+CH2CHO | 7.00E+14 | 0 | 42000 | Same reactivity assumed as C2H5O2H (Konnov, 2000) |
| 144 | C2H3OOH+O=OH+C2H3O2 | 2.80E+13 | 0 | 6400 | Same reactivity assumed as C2H5O2H |
| 145 | C2H3OOH+O=OH+CHCHOOH | 6.00E+13 | 0 | 10000 | Same reactivity assumed as C2H5O2H |
| 146 | C2H3OOH+OH=C2H3O2+H2O | 6.00E+11 | 0 | -380 | Same reactivity assumed as C2H5O2H |
| 147 | C2H3OOH+OH=CHCHOOH+H2O | 6.00E+11 | 0 | -380 | Same reactivity assumed as C2H5O2H |
| 148 | C2H3OOH+C2H=C2H3O2+C2H2 | 6.00E+12 | 0 | 0 | Same reactivity assumed as C2H5O2H |
| 149 | CH3COOO+CH3OO= CH3CO2H+HCHO+O2 | 2.47E+09 | 0 | -4200 | Moortgat et al., 1989 |
| 150 | CH3CO+OH=CH3CO2H | 1.00E+14 | 0 | 0 | See present work |
| 151 | CH3CO2H=CH4+CO2 | 7.08E+13 | 0 | 74600 | Duan and Page, 1995 |
| 152 | CH3CO2H=CH2CO+H2O | 4.47E+14 | 0 | 79800 | Duan and Page, 1995 |
| 153 | CH3CO2H+OH=CH3CO2+H2O | 2.40E+11 | 0 | -400 | DeMore et al., 1997 |
| 154 | CH2OH+CHO=HOCH2CHO | 1.80E+13 | 0 | 0 | Estimate |
| 155 | HOCH2CHO+OH=HOCH2CO+H2O | 6.00E+12 | 0 | 0 | Atkinson et al., 2005 |
| 156 | HOCH2CHO+OH=HOCHCHO+H2O | 1.50E+12 | 0 | 0 | Atkinson et al., 2005 |
| 157 | C2H5+O2=C2H4OOH | 1.02E+50 | -12.4 | 15460 | Thompson et al., 1994 |
| 158 | C2H5OOH+O=OH+C2H4OOH | 3.55E+06 | 2.4 | 5830 | Same rate constant as C2H6 + O (Konnov, 2000) |
| 159 | C2H5OOH+OH=C2H4OOH+H2O | 6.00E+11 | 0 | -380 | Baulch et al., 2005 |
| 160 | C2H5OOH+O=OH+CH3CHOOH | 6.60E+13 | 0 | 4150 | Baulch et al., 2005 (total rate attributed to this channel) |
| 161 | C2H5OOH+OH=CH3CHOOH+H2O | 6.00E+11 | 0 | -380 | Same reactivity assumed as C2H4OOH |
| 162 | CH3CHOOH=CH3CHO+OH | 3.16E+11 | 0 | 19500 | Estimate |
| 163 | CH3CHOOH=C2H4O#3+OH | 3.16E+11 | 0 | 19500 | Estimate |
| | **$C_3$ species** | | | | |
| 164 | CHO+C2H5=>C2H5CHO | 1.80E+13 | 0 | 0 | Tsang and Hampson, 1986 |
| 165 | C2H5CHO+H=H2+C2H5CO | 4.00E+13 | 0 | 4200 | Same reactivity assumed as CH3CHO (Barbé et al., 1995) |
| 166 | C2H5CHO+CH3=C2H5CO+CH4 | 2.00E-06 | 5.6 | 2500 | Same reactivity assumed as CH3CHO |
| 167 | C2H5CHO+C2H3=C2H4+C2H5CO | 8.10E+10 | 0 | 3700 | Same reactivity assumed as CH3CHO |
| 168 | C2H5CHO+C2H5=C2H6+C2H5CO | 1.30E+12 | 0 | 8500 | Same reactivity assumed as CH3CHO |
| 169 | C2H5CHO+O=C2H5CO+OH | 1.40E+13 | 0 | 2300 | Same reactivity assumed as CH3CHO |
| 170 | C2H5CHO+OH=C2H5CO+H2O | 4.20E+12 | 0 | 500 | Same reactivity assumed as CH3CHO |
| 171 | C2H5CHO+CH3O=C2H5CO+CH3OH | 2.40E+11 | 0 | 1800 | Same reactivity assumed as CH3CHO |
| 172 | CH2CO+CH3=C2H5CO | 2.40E+12 | 0 | 8000 | Borisov et al., 1990 Formation of C2H5CO assumed instead of that of C2H5+CO |
| 173 | C2H5CO+OH=C2H5CO2H | 1.00E+14 | 0 | 0 | See present work |
| 174 | C2H5CO=C2H5+CO | 5.89E+12 | 0 | 14400 | Watkins and Thompson, 1973 |
| 175 | C2H5CO+H=C2H5+CHO | 9.60E+13 | 0 | 0 | Same reactivity assumed as CH3CO (Barbé et al., 1995) |
| 176 | C2H5CO+B6CH2=C2H5+CH2CO | 1.80E+13 | 0 | 0 | Same reactivity assumed as CH3CO |
| 177 | C2H5CO+B5CH2=C2H5+CH2CO | 1.80E+13 | 0 | 0 | Same reactivity assumed as CH3CO |
| 178 | C2H5CO+O=C2H5+CO2 | 9.60E+12 | 0 | 0 | Same reactivity assumed as CH3CO |
| 179 | C2H5CO+OH=>C2H5+CO+OH | 3.00E+13 | 0 | 0 | Same reactivity assumed as CH3CO |
| 180 | C2H5CO+CHO=C2H5CHO+CO | 9.00E+12 | 0 | 0 | Same reactivity assumed as CH3CO |
| 181 | C2H5CO+HCHO=C2H5CHO+CHO | 1.80E+11 | 0 | 12900 | Same reactivity assumed as CH3CO |
| 182 | C2H5CO+CH3O=HCHO+C2H5CHO | 6.00E+12 | 0 | 0 | Same reactivity assumed as CH3CO |
| 183 | C2H5CO+CH3OH=C2H5CHO+CH2OH | 4.85E+03 | 3.0 | 12300 | Same reactivity assumed as CH3CO |
| 184 | O2+C2H5CO=C2H5COOO | 2.40E+12 | 0 | 0 | Same reactivity assumed as CH3CO |
| 185 | C2H5CO2H=C2H6+CO2 | 7.08E+13 | 0 | 74600 | See present work |
| 186 | C2H5CO2H+OH=C2H5CO2+H2O | 2.40E+11 | 0 | -400 | See present work |
| 187 | C2H5CO2+M=C2H5+CO2+M | 8.70E+15 | 0 | 14400 | See present work |
| 188 | C2H5COOO+CH3OO= C2H5CO2H+HCHO+O2 | 2.47E+09 | 0 | -4200 | Same reactivity assumed as CH3COOO (Barbé et al., 1995) |
| 189 | C2H5COOO+C2H4O#3= C2H5COOOH+CH2CHO | 1.00E+12 | 0 | 9300 | Same reactivity assumed as CH3COOO |
| 190 | C2H5COOO+HO2= C2H5COOOH+O2 | 5.50E+10 | 0 | -2600 | Same reactivity assumed as CH3COOO |
| 191 | C2H5COOO+C2H5OOH= C2H5COOOH+C2H5OO | 5.00E+11 | 0 | 9200 | Same reactivity assumed as CH3COOO |
| 192 | C2H5COOO+C2H5COOO=> 2C2H5+O2+2CO2 | 1.70E+12 | 0 | -1000 | Same reactivity assumed as CH3COOO |
| 193 | C2H5COOOH=>C2H5+CO2+OH | 1.00E+16 | 0 | 40000 | Same reactivity assumed as CH3COOO |
| 194 | CH2CHCO+OH=C2H3CO2H | 1.00E+14 | 0 | 0 | See present work |
| 195 | C2H3CO2H=C2H4+CO2 | 7.08E+13 | 0 | 74600 | See present work |
| 196 | C2H3CO2H+OH=C2H3CO2+H2O | 2.40E+11 | 0 | -400 | See present work |
| 197 | C2H3CO2+M=C2H3+CO2+M | 8.70E+15 | 0 | 14400 | See present work |

**Figures captions:**

Figure 1: Modeling of a laminar premixed flame of propane at 1 atm for an equivalence ratio of 0.9. Symbols correspond to experiments of Biet et al. (2005) and lines to simulations.

Figure 2: Modeling of a laminar premixed flame of propane at 1 atm for an equivalence ratio of 0.48. Symbols correspond to experiments of Biet et al. (2005) and lines to simulations.

Figure 3: Modeling of the formation of acids in laminar premixed flames of propane at 1 atm for an equivalence ratio of (a) 0.9 and (b) 0.48.

Figure 4: Comparison between experimental (Zervas (2005)) and simulated proportions of acids for an equivalence ratio of 0.9 at the maximum of the peak of formic acid.

Figure 5; Main pathways of formation of the four monocarboxylic acids studied under the conditions of figure 3a for a distance above the burner corresponding to the maximum of the peak of formic acid (0.5 cm). The size of the arrows is proportional to the relative flow rates.

Figure 6: Simulated profiles of the main aromatic species under the conditions of the flame of propane of Biet et al. (2005) at $\phi$=0.9, but with an addition of 10% toluene (relative to the mole fraction of propane).

Figure 7: Formation of acids under the conditions of the flame of propane of Biet et al. (2005) at $\phi$=0.9, but with an addition of 10% toluene (relative to the mole fraction of propane).

Figure 8: Simulated evolution of (a) the main $C_1$-$C_3$, (b) $C_7$ and $C_8$ oxygenated species during the pre-ignition phase in an HCCI engine.

Figure 9: Comparison of the mole fractions of acetaldehyde, acrolein, propionaldehyde and methanol relative to that of formaldehyde computed in this study and measured by Zervas et al. (2002).

Figure 10: Formation of $C_1$-$C_3$ acids and formaldehyde according to the different proposed models during the pre-ignition phase in an HCCI engine.

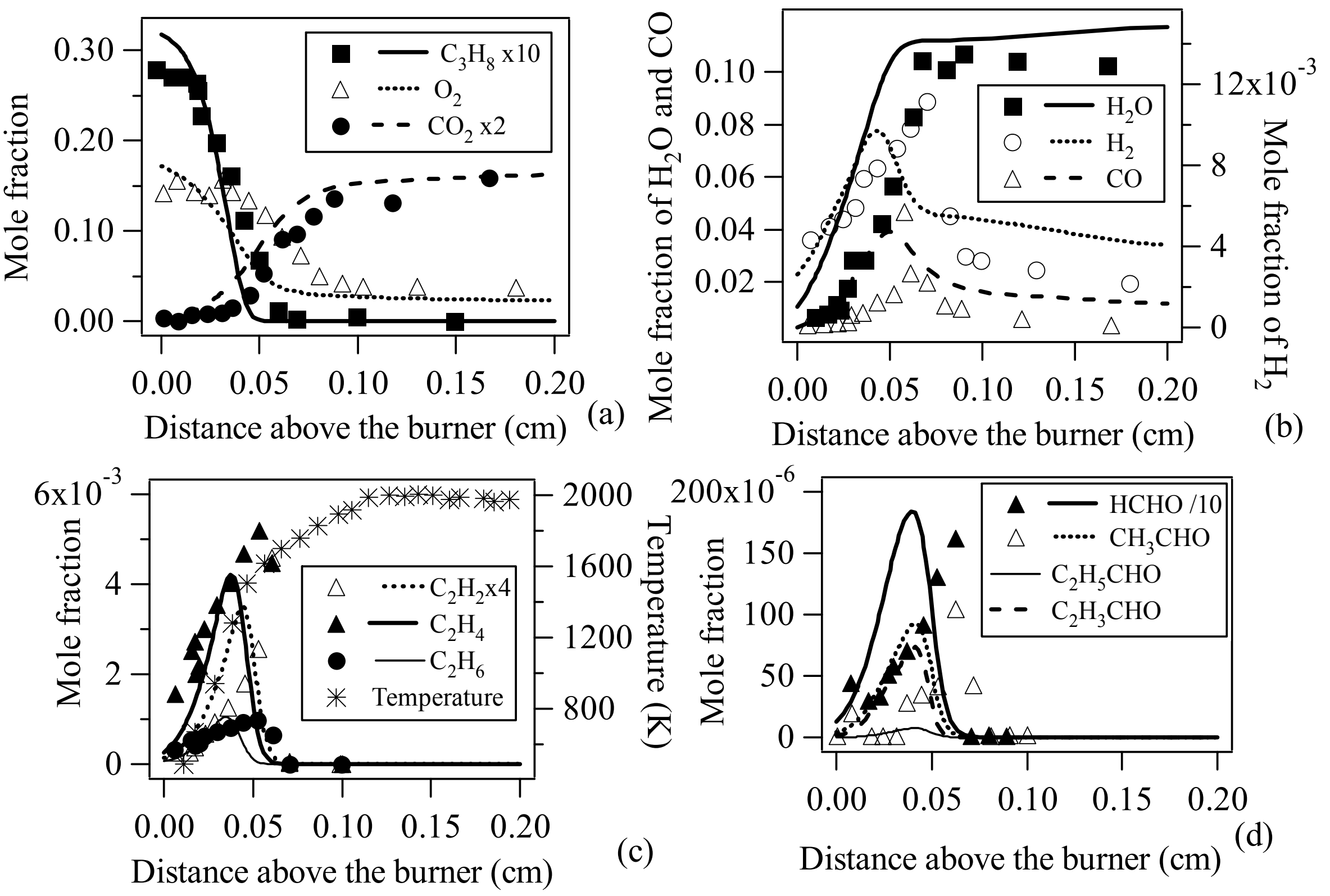

Figure 2

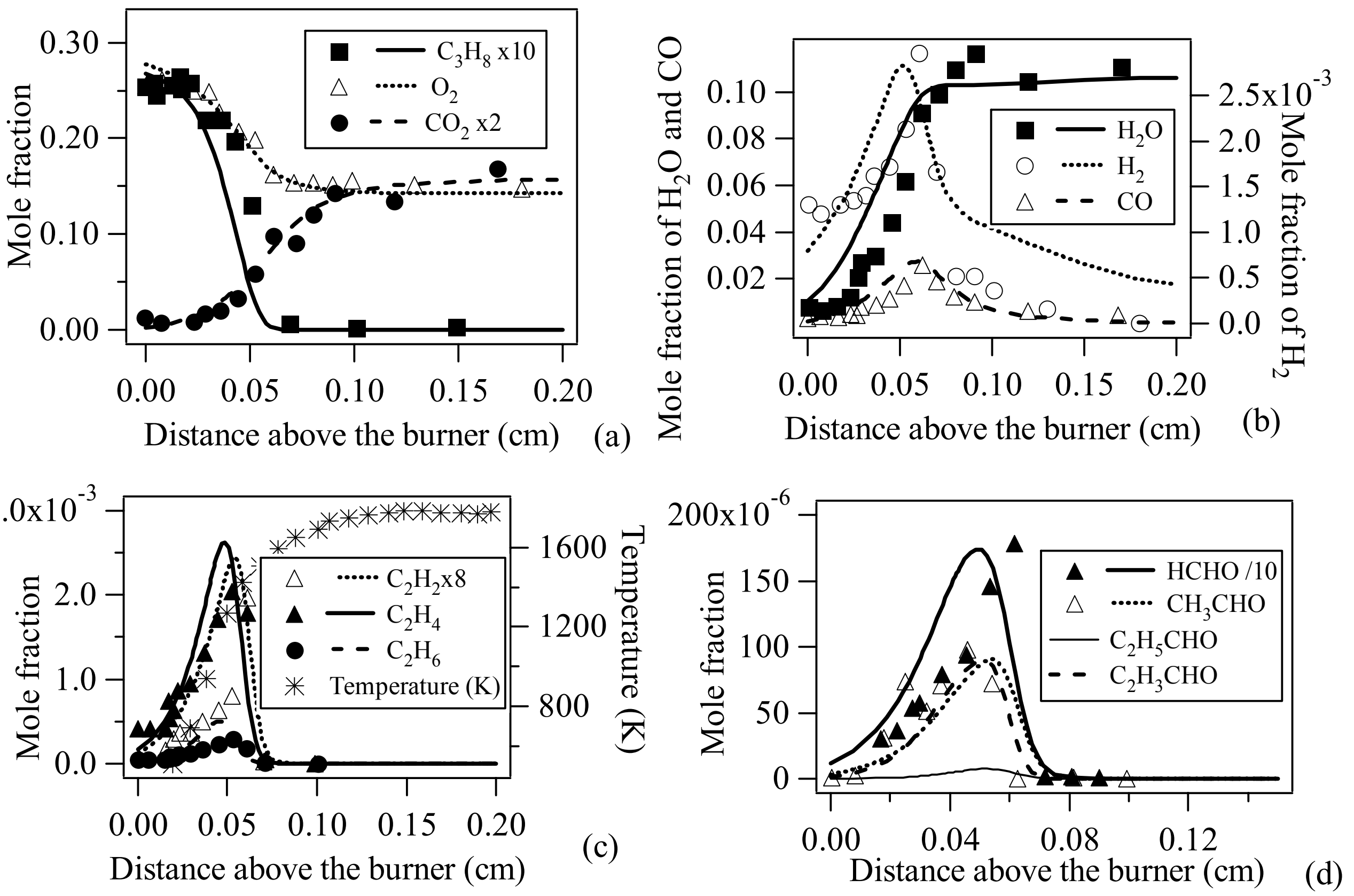

C3H8 x10
O2
CO2 x2
Mole fraction
Distance above the burner (cm)
(a)
H2O
H2
CO
Mole fraction of H2O and CO
Mole fraction of H2
Distance above the burner (cm)
(b)
C2H2x8
C2H4
C2H6
Temperature (K)
Mole fraction
Temperature (K)
Distance above the burner (cm)
(c)
HCHO /10
CH3CHO
C2H5CHO
C2H3CHO
Mole fraction
Distance above the burner (cm)
(d)

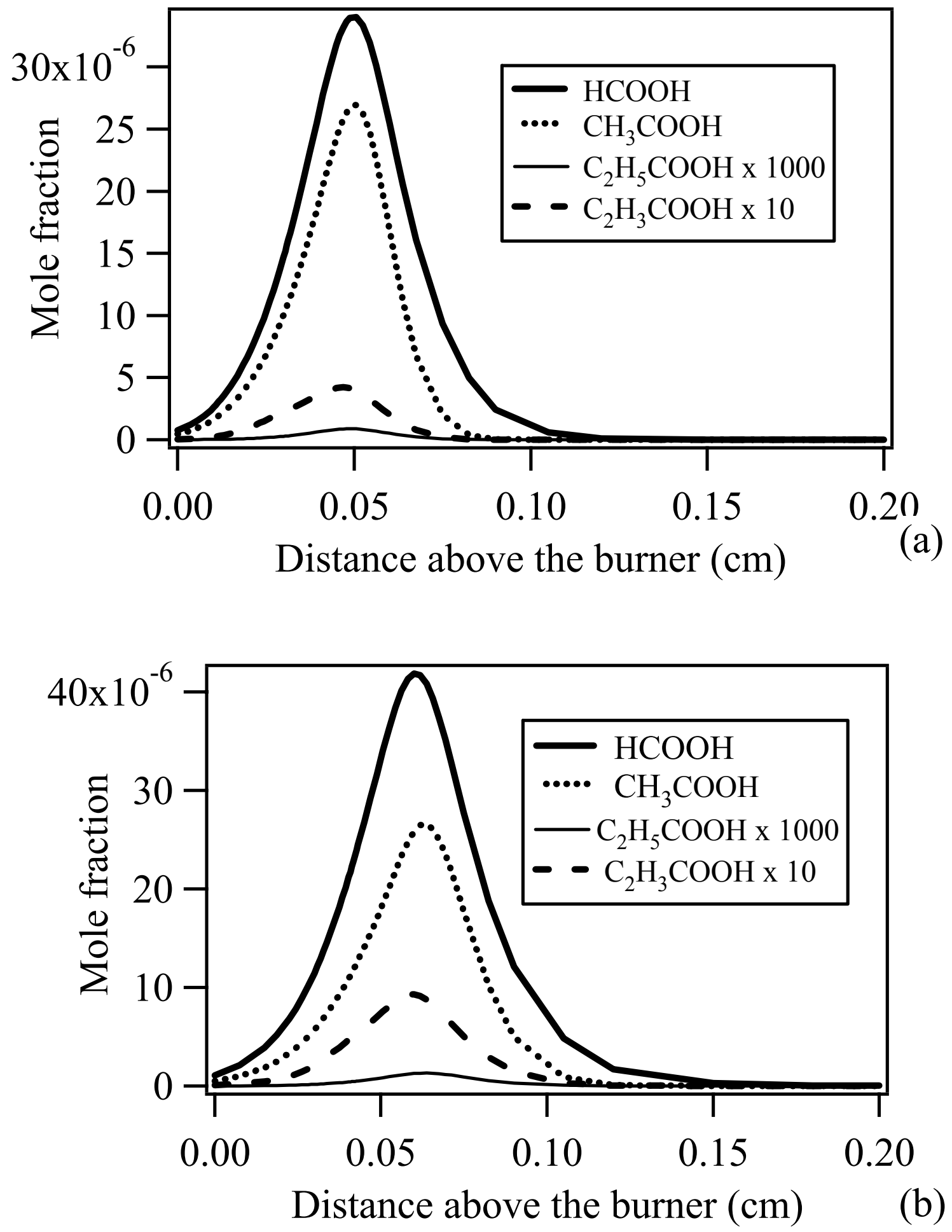

Figure 4

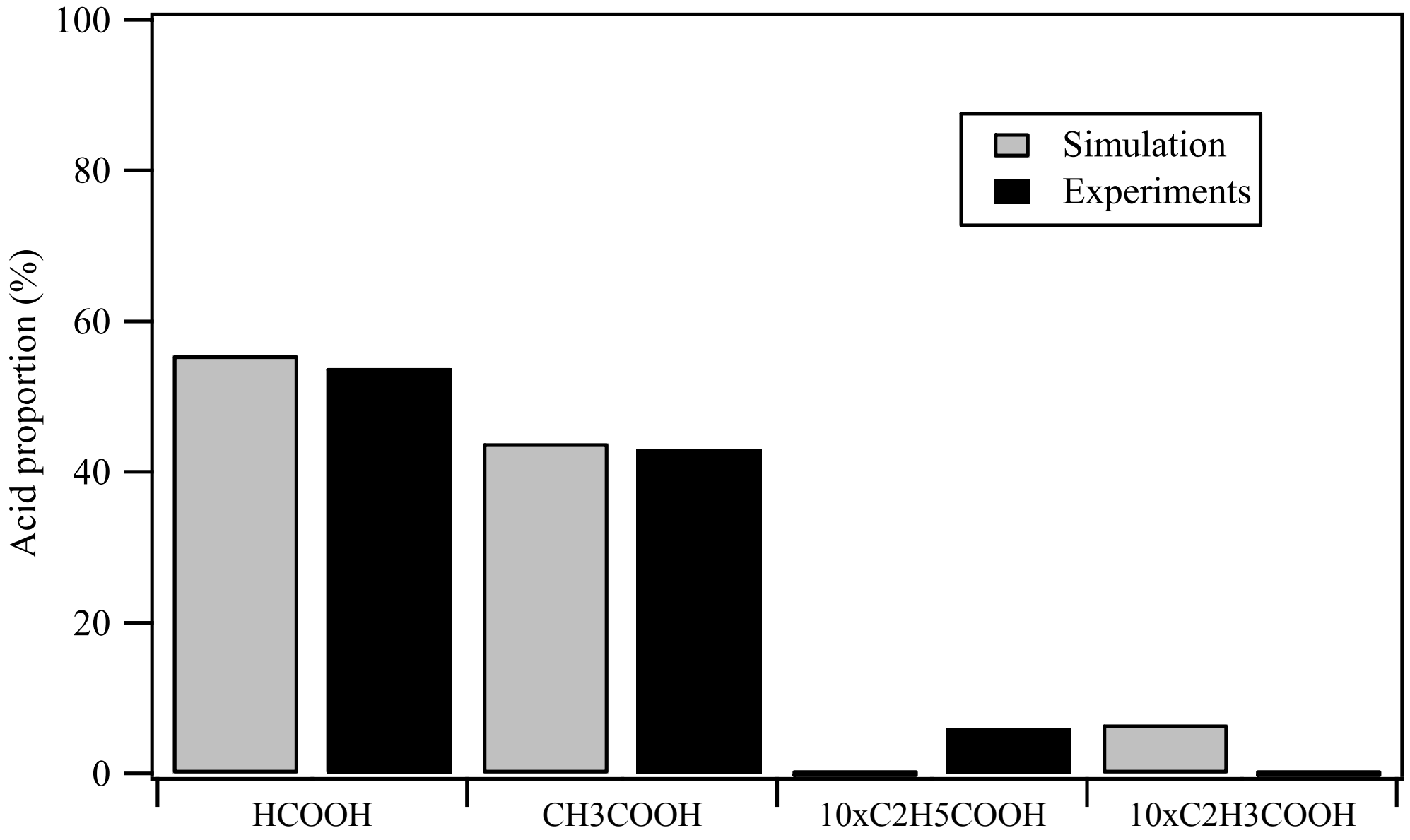

Figure 5

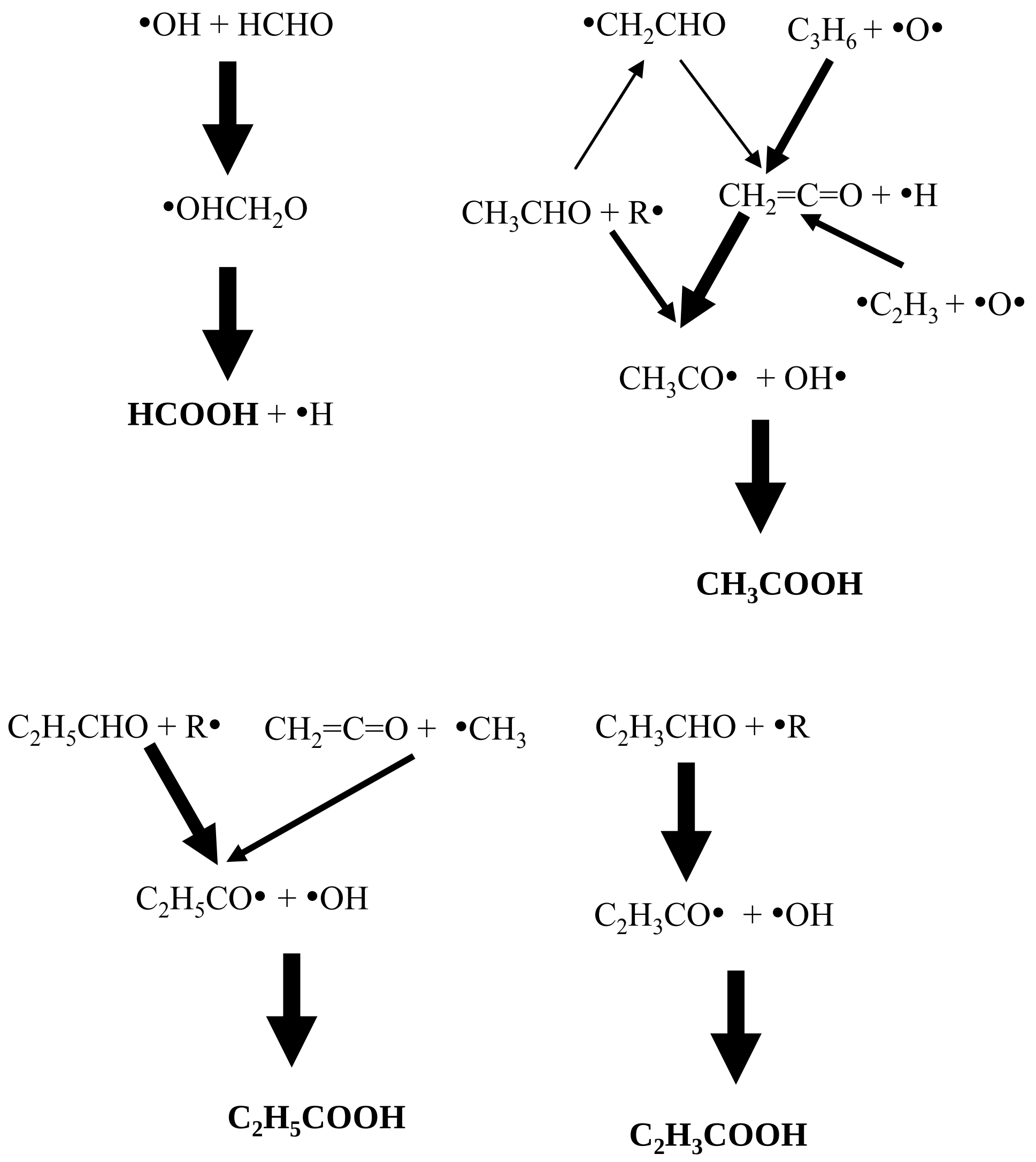

Figure 6

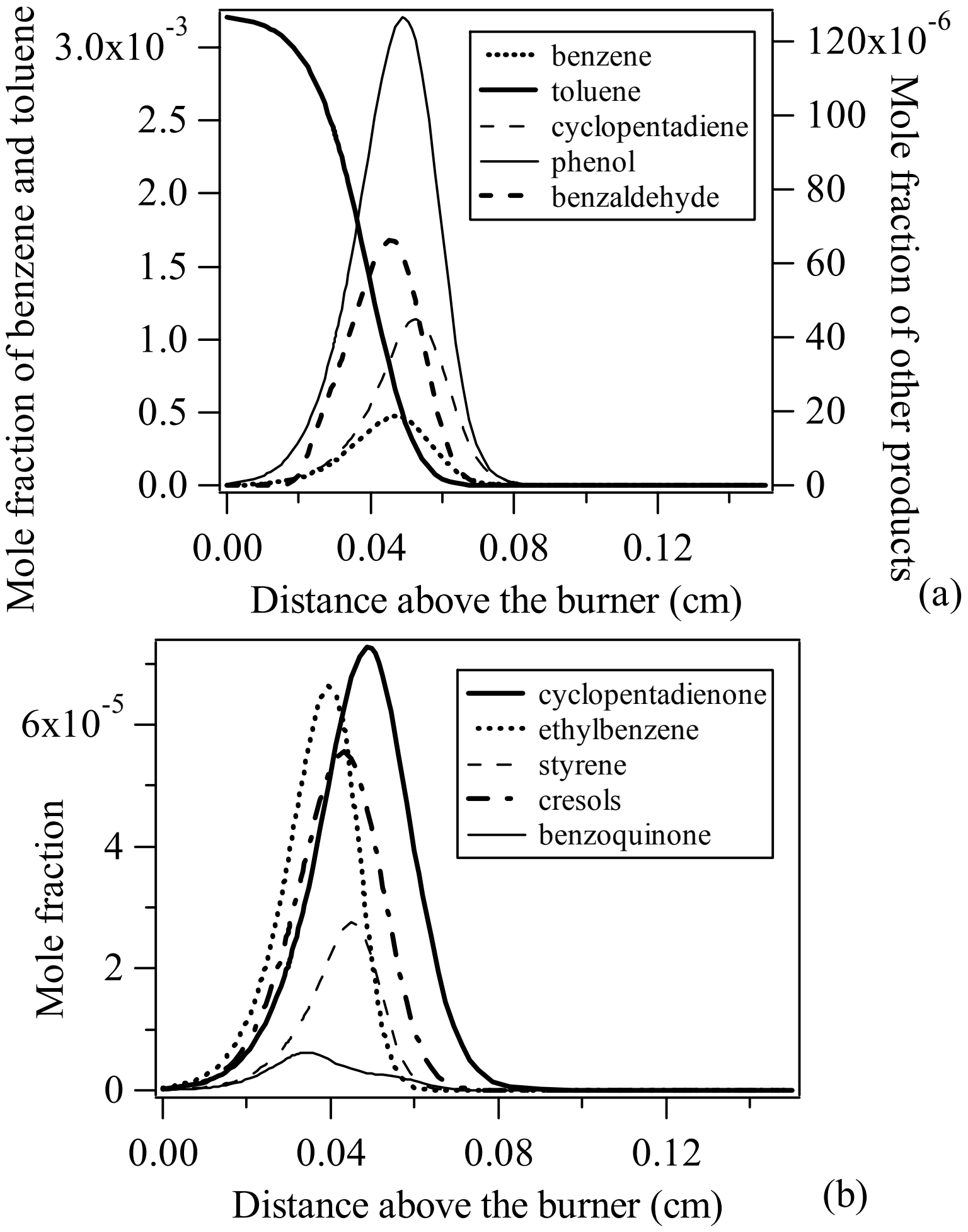

Figure 7

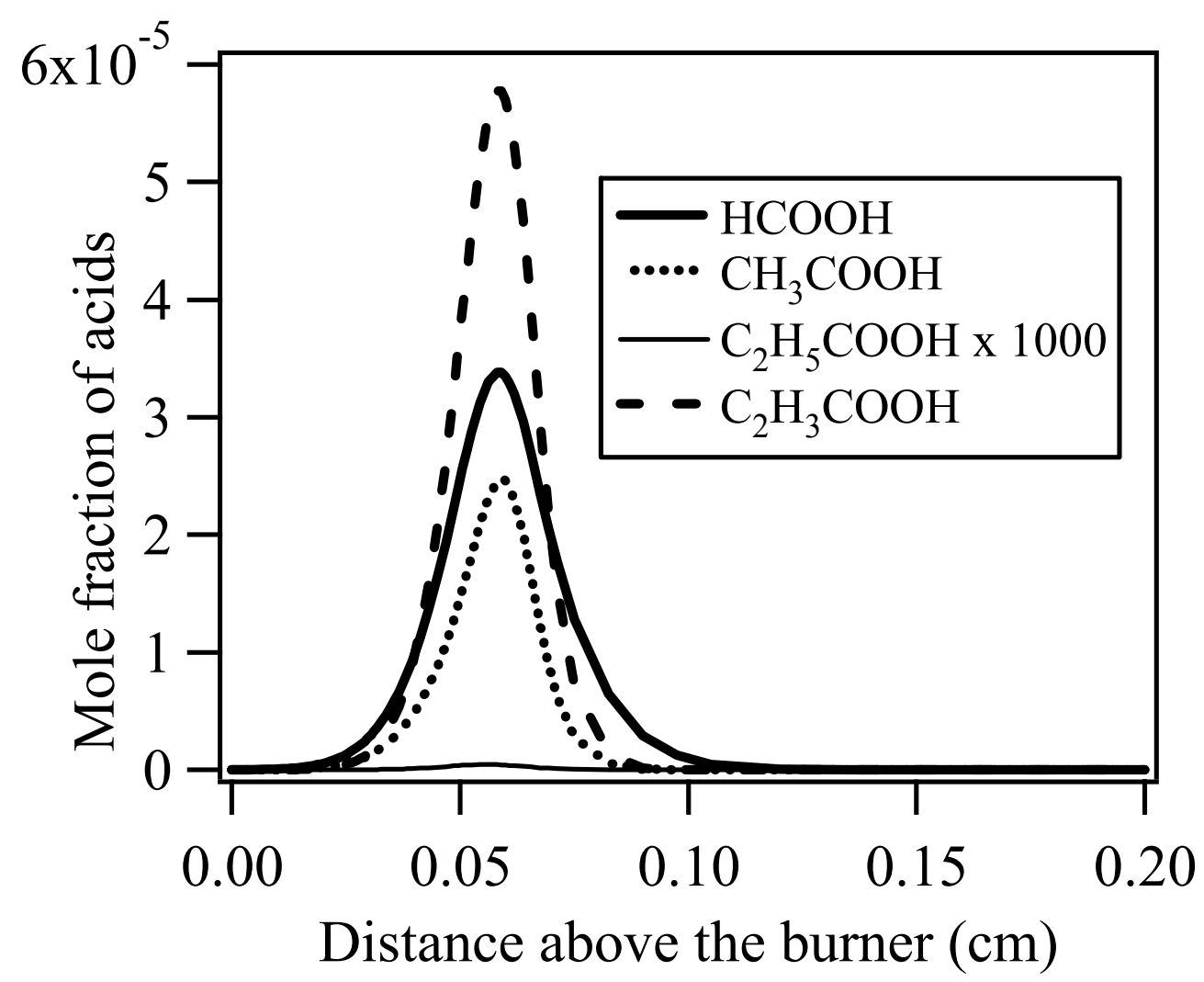

Figure 8

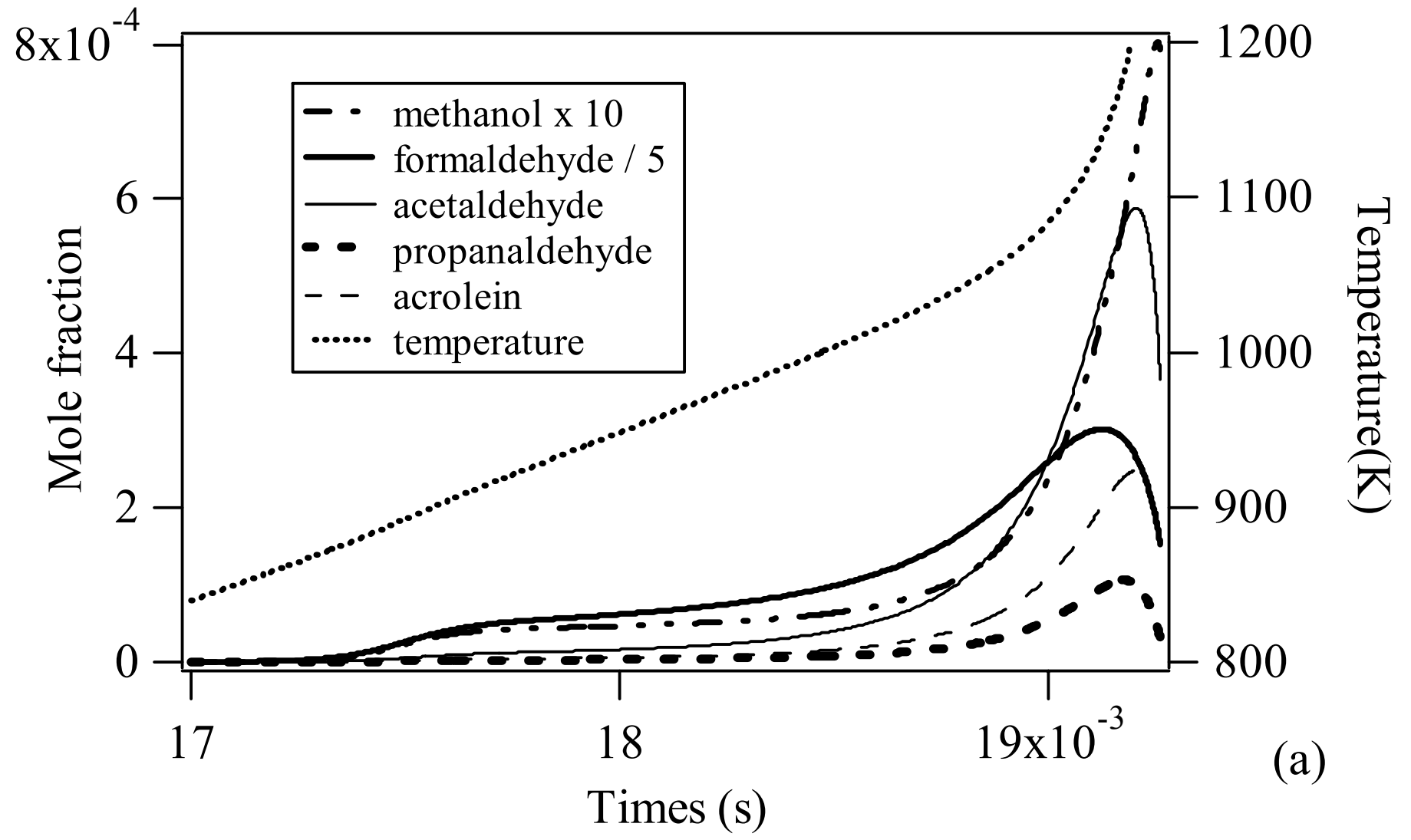

(a)

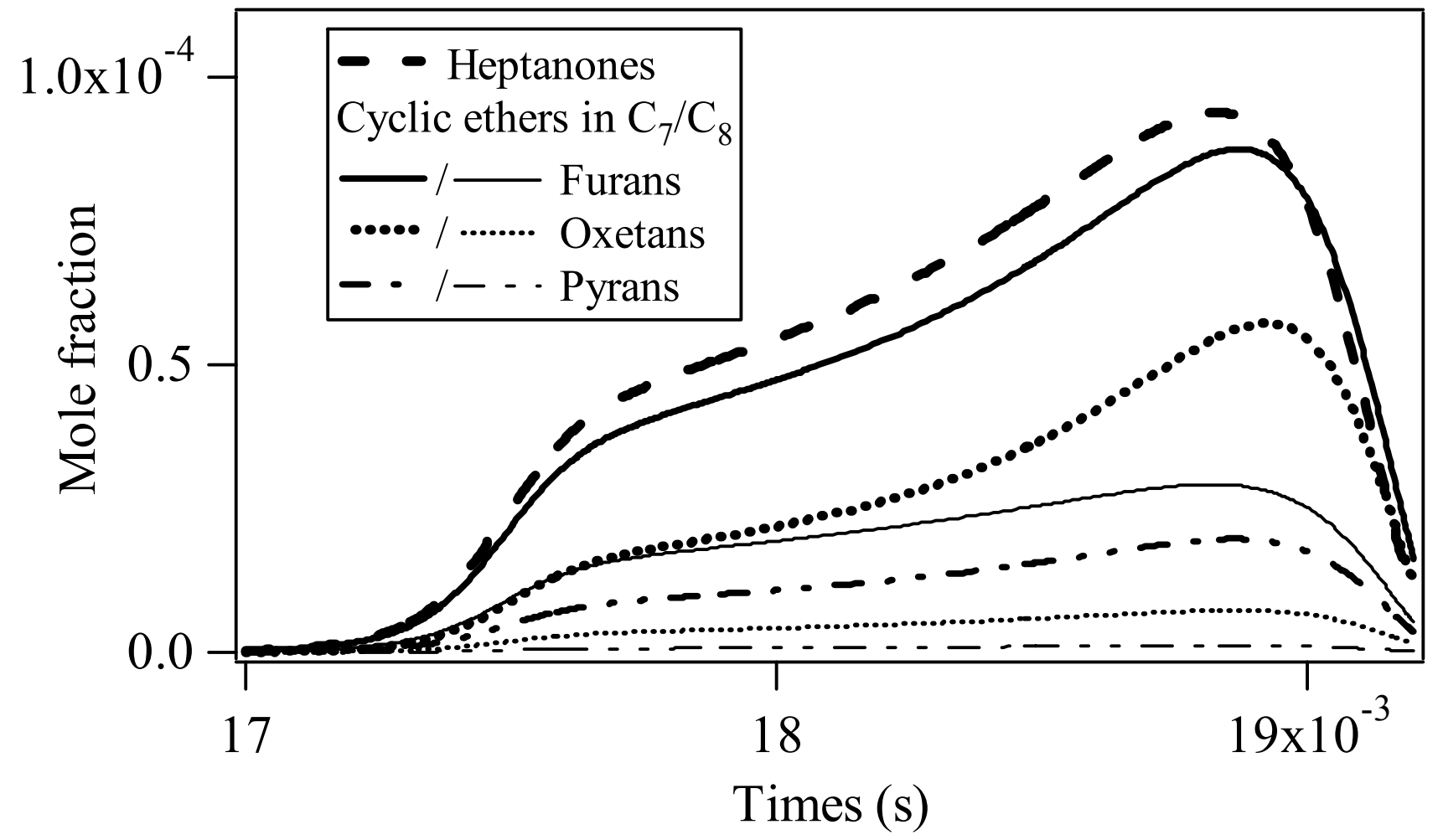

(b)

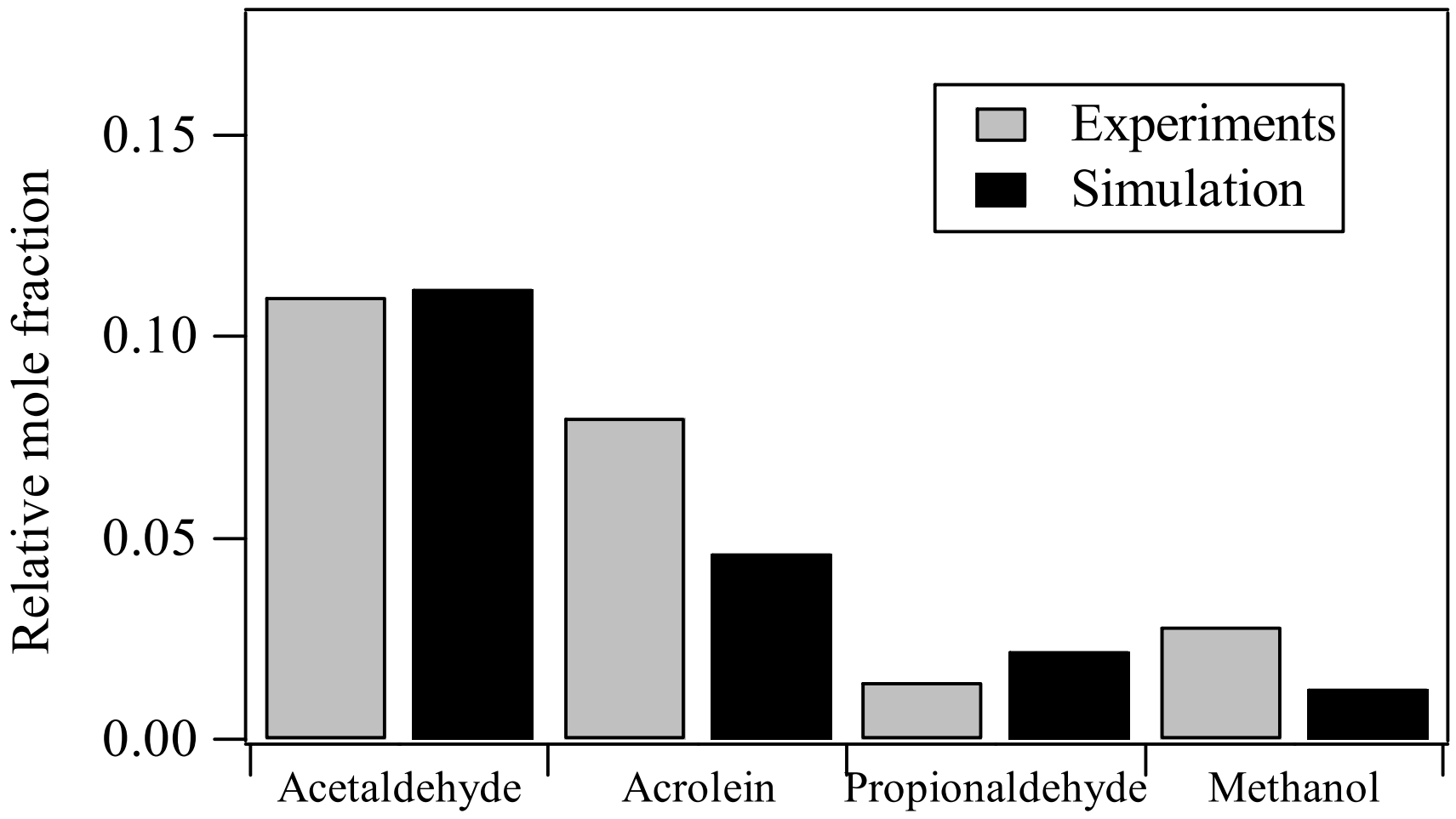

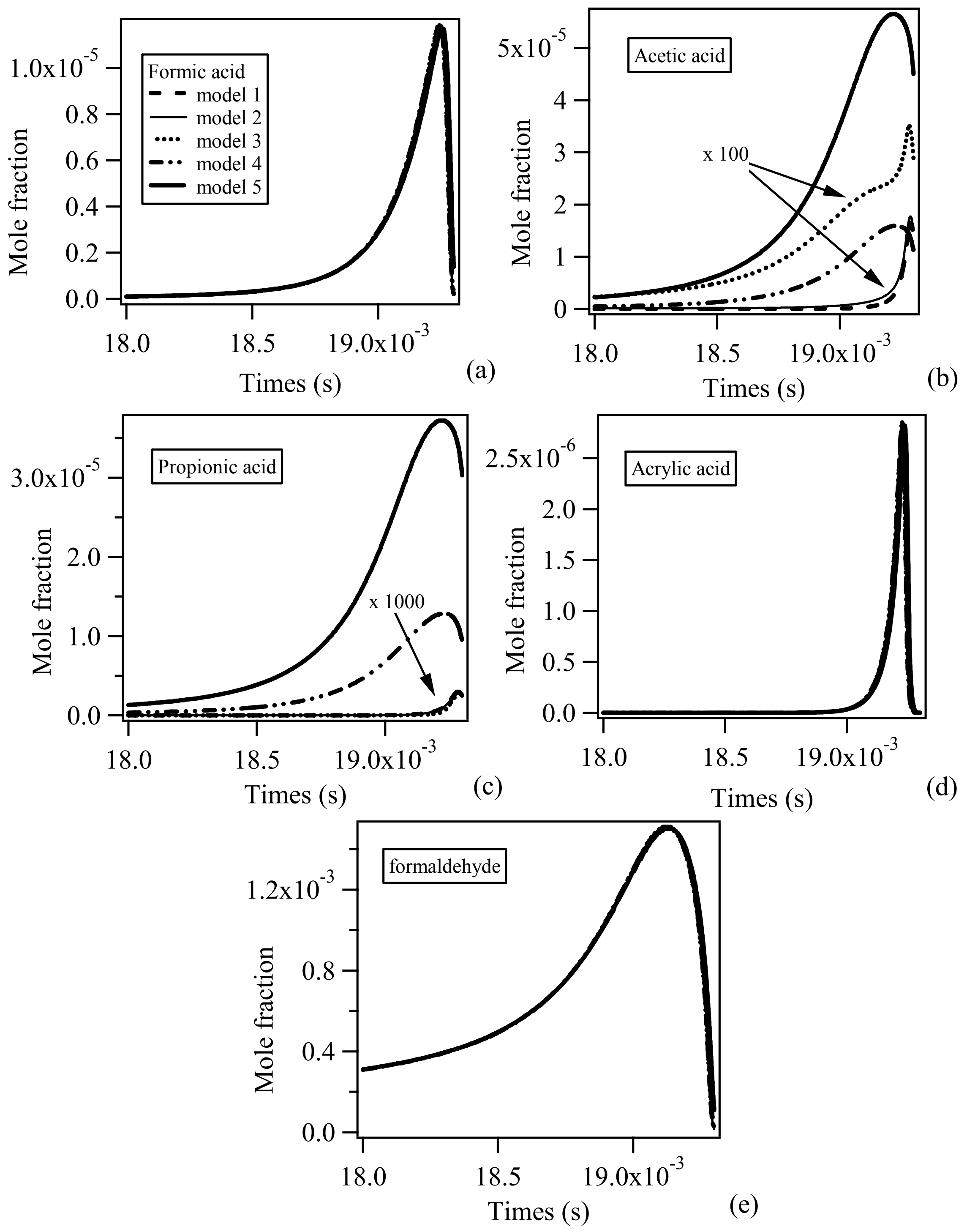